\documentclass[letterpaper]{article} 
\usepackage{aaai2027}  
\usepackage[hyphens]{url}  
\usepackage{graphicx} 
\usepackage{natbib}  
\usepackage{caption} 
\usepackage{algorithm}
\usepackage{algorithmic}
\usepackage{amsmath}
\usepackage{amssymb}
\usepackage{bm}
\usepackage{newfloat}
\usepackage{listings}
\DeclareCaptionStyle{ruled}{labelfont=normalfont,labelsep=colon,strut=off} 
\floatstyle{ruled}
\newfloat{listing}{tb}{lst}{}
\floatname{listing}{Listing}

\usepackage{booktabs}
\usepackage{multirow}

\nocopyright 

\title{Local-to-Global Sentence-Level Graph Reranking for Scientific Synthesis}
\author{
    Zheng Dou\textsuperscript{\rm 1},
    Zhao Zhang\textsuperscript{\rm 1}\corresponding,
    Hao Geng\textsuperscript{\rm 1},
    Ningjing Wang\textsuperscript{\rm 1},
    Deqing Wang\textsuperscript{\rm 1,2}\corresponding
}
\affiliations{
    \textsuperscript{\rm 1} School of Computer Science and Engineering, Beihang University, Beijing, China\\
    \textsuperscript{\rm 2} Zhongguancun Laboratory, Beijing, China\\ 

    \{miracle\_dz, zhao\_zhang, genghao, wangningjing, dqwang\}@buaa.edu.cn
}

\begin{document}

\maketitle

\begin{abstract}
Retrieval-augmented scientific synthesis aims to answer complex research questions by integrating information from multiple papers into comprehensive and well-grounded responses. Since the generator can only synthesize the information selected and organized by the reranker, the quality of the generated synthesis depends critically on the reranked results. However, most rerankers operate at the passage level, which leaves key methodological, empirical, and comparative information buried in long and flat contexts, weakening the grounding of generated claims. Moreover, existing rerankers mainly rely on independent query-candidate scoring which overlooks complementary, contextual, and contrasting relations across scientific candidates, limiting information coverage and the comprehensiveness of the resulting synthesis. To address these limitations, we propose LoG-Reranker, a local-to-global sentence-level graph reranking framework for scientific synthesis. LoG-Reranker performs role-aware local scoring to identify fine-grained, query-relevant sentences and then models their relations on a sentence graph across the candidate set to globally refine sentence rankings. Top-ranked sentences and their connected neighbors are organized into a structured input context for generator to produce more grounded and comprehensive synthesis. Extensive experiments on scientific synthesis and reranking benchmarks show that LoG-Reranker consistently outperforms competitive rerankers, yielding more reliable rankings and improving the quality of generated synthesis.
\end{abstract}

\section{Introduction}
\begin{figure}[t]
  \centering
  \includegraphics[width=\columnwidth]{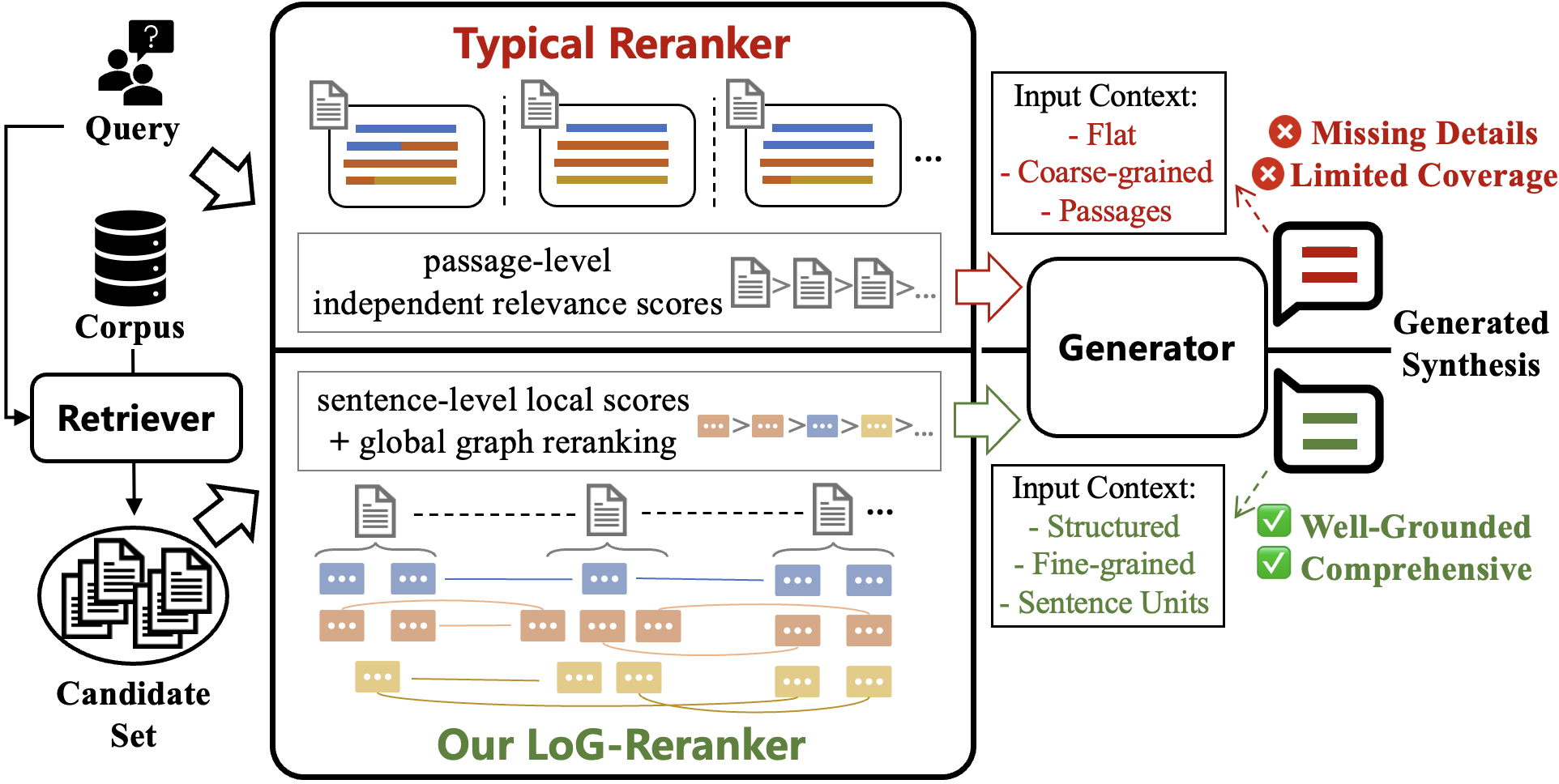}
  \caption{The pipeline of scientific synthesis and the motivation of LoG-Reranker.}
  \label{fig:intro}
\end{figure}
With recent advances in LLMs and retrieval augmentation, RAG has become a natural approach to scientific question answering (QA) by grounding LLM responses in external papers and passages \cite{SciRAG,LitLLMs}.
Unlike typical scientific QA tasks that seek short or isolated answers, \textit{scientific synthesis} aims to generate long-form, comprehensive and well-grounded answers to complex research questions by integrating information from multiple scientific papers \cite{OpenScholar}.
With the explosive growth of scientific literature, this task has become increasingly essential for supporting researchers in literature review and research discovery.
Figure~\ref{fig:intro} shows a typical pipeline for scientific synthesis: a retriever first recalls a candidate set of potentially relevant passages from a large corpus, then a reranker scores these candidates via query-candidate interaction to prioritize more relevant information, and a generator synthesizes the top-ranked candidates into a final answer.
This paper focuses on the reranking stage, which serves as the bridge between retrieval and generation. By selecting and organizing the source information supplied to the generator, an effective reranker should provide higher-quality input context and thereby enable more \textit{grounded} and \textit{comprehensive} scientific synthesis.

However, typical rerankers still operate at the document or passage level \cite{BGE,GTE}, remaining limited for identifying key methodological, empirical, and comparative information within detailed sentences of scientific candidates.
Such coarse-grained selection yields long and flat contexts for the generator, where critical details may be buried \cite{Lost}. Without these details, the generator may not reliably identify and attribute the source content to support generated claims \cite{ReClaim,LongCite}, resulting in less grounded synthesis.
This highlights the need for reranking that moves beyond coarse passage-level prioritization to surface finer-grained information units, such as sentences, before synthesis generation.

Another limitation is that typical rerankers score each query-candidate pair independently \cite{CrossEncoder,RankT5}, overlooking how candidates complement, contextualize, or contrast with one another. 
This limits the generator’s ability to ensure coherent integration and sufficient coverage across the relevant aspects of a research question \cite{LongFormQAEval,ICAT}, leading to less comprehensive synthesis.
Although some graph-based rerankers introduce relational signals through passage similarity \cite{GraphER}, document links \cite{GNRR}, or entity-level knowledge graphs \cite{KERM}, these structures still fail to capture fine-grained relations among methodological, empirical, and comparative statements across scientific candidates. 
Therefore, an effective reranker for scientific synthesis should consider both local query relevance and the global relational structure within the candidate set.

Motivated by these limitations, we propose LoG-Reranker, a local-to-global sentence-level graph reranking framework for scientific synthesis. 
As illustrated in Figure~\ref{fig:intro}, our reranker replaces coarse, independently ranked passages with fine-grained sentence selection and cross-candidate relational modeling. 
The local component identifies query-relevant details, while the global component promotes complementary and contextually connected information across candidates. 
This design treats reranking as synthesis-oriented context refinement rather than only relevance ordering. 
Together, they provide the generator with structured sentence-level context, addressing missing details and limited coverage to generate more grounded and comprehensive scientific synthesis.
The main contributions of this work are summarized as follows:
\begin{itemize}
  \item We introduce role-aware local sentence scoring that models the compatibility between query intents and sentence roles for fine-grained relevance estimation.
  \item Building on these local relevance scores, we propose a local-to-global graph reranking framework that captures intra-passage adjacency and constrained cross-passage relations in a sentence graph to globally refine sentence rankings over the retrieved candidate set.
  \item Extensive experiments on scientific synthesis and reranking benchmarks show that LoG-Reranker improves both reranking and synthesis generation, including citation grounding, information coverage, and answer quality.
\end{itemize}

\section{Related Work}
\paragraph{Neural and LLM-Based Reranking.}
Neural rerankers initially model query-candidate relevance through pretrained cross-encoders \cite{CrossEncoder}, while subsequent generative approaches reformulate relevance estimation with sequence-to-sequence models \cite{SequenceToSequence} and ranking-specific objectives \cite{RankT5}.
Recent listwise methods further incorporate inter-candidate comparisons through multi-candidate fusion \cite{ListT5} or LLM-based permutation generation \cite{ZeroShot,RankGPT,RankZephyr}.
In scientific domain, DeepEra introduces step-by-step reasoning into scientific reranking to assess logical relevance and evidential support beyond surface-level semantic similarity \cite{DeepEra}.
Nevertheless, these methods remain primarily relevance-oriented and operate at the document or passage level, without explicitly modeling fine-grained scientific information for downstream generation.
\paragraph{Fine-Grained Reranking and Information Refinement.}
To reduce irrelevant content in coarse-grained candidates, recent methods refine retrieved information through extractive or abstractive compression before generation \cite{Recomp}.
Moving beyond compression, sentence-level reranking and reconstruction select relevant content within passages rather than treating each passage as an indivisible unit \cite{DSLR}.
Subsequent approaches improve such refinement by preserving contextual dependencies \cite{EXIT}, prioritizing evidential support \cite{ECoRAG}, or identifying fine-grained knowledge overlooked by document-level processing \cite{Fine-grained}.
In scientific reranking, compact document features have also been used in a coarse-to-fine process to enlarge the candidate pool before refining the highest-ranked documents \cite{CoRank}.
Despite exposing more concise and informative content, these methods mainly compress, select, or rerank fine-grained content without globally modeling their relations across the candidate set.
\paragraph{Graph-Based Reranking.}
To move beyond isolated query-candidate scoring, graph-based rerankers \cite{dont} model relations among candidates through corpus-level connectivity and semantic proximity, enabling adaptive candidate expansion \cite{GAR} and relational score propagation \cite{GNRR}.
Related approaches further enrich passage relevance by incorporating explicit entity and relation knowledge into representations \cite{KERM}, or by constructing candidate graphs from heterogeneous proximity signals to refine their ranking scores \cite{GraphER}.
To capture relations among the fine-grained information units, recent studies construct graphs at the sentence level.
Sentence graphs have been used to jointly model question-answer sentences for answer selection \cite{Joint}, connect relevant sentences during progressive multi-hop reasoning \cite{ChainRAG}, and organize sentence-level dependencies for graph-guided selection and path expansion \cite{SentGraph}.
However, these methods are primarily designed for answer selection or multi-hop reasoning, whereas scientific synthesis requires sentence-level graph reranking to assess fine-grained scientific statements within their broader cross-paper context.
We propose LoG-Reranker to fill this gap by integrating query-aware local sentence scoring with graph-based global reranking over the retrieved candidate set for scientific synthesis.

\section{Method}
\begin{figure*}[t!]
  \centering
  \includegraphics[width=0.96\textwidth]{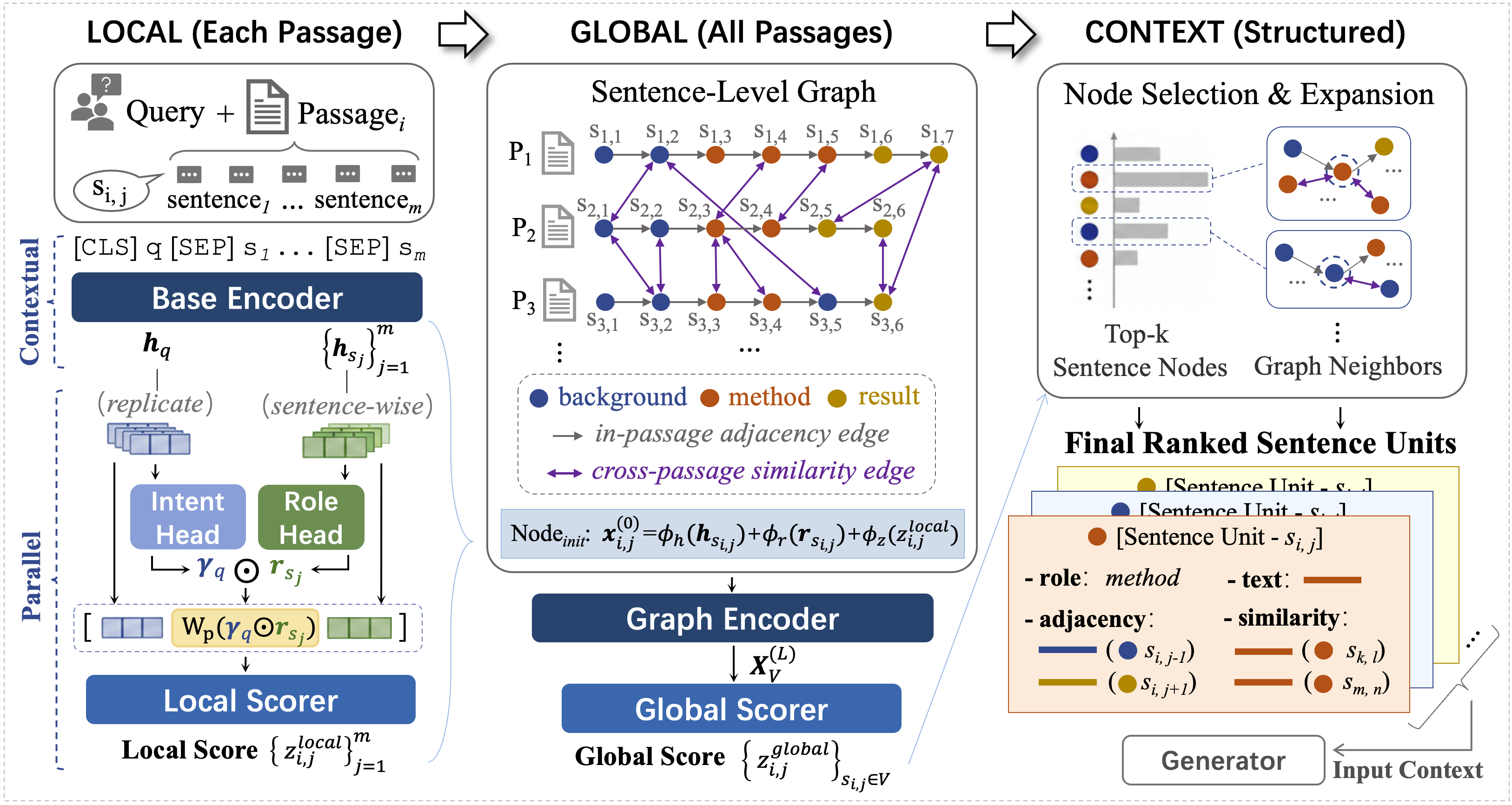}
  \caption{The overall framework of LoG-Reranker.}
  \label{fig:method}
\end{figure*}
\subsection{Problem Formulation}
Given a scientific query $q$ and a corpus $\mathcal{C}$, a retriever first returns a candidate set of $N$ potentially relevant passages $\mathcal{P}_q=\{p_1,\ldots,p_N\}.$
Reranking operates over the retrieved set $\mathcal{P}_q$ to provide more relevant information for downstream generation.
Each retrieved passage $p_i\in\mathcal{P}_q$ is further decomposed into sentences $p_i=\{s_{i,1},\ldots,s_{i,m_i}\}$, together yielding a unified sentence candidate set $\mathcal{S}_q$.
The goal of sentence-level reranking is to learn a scoring function $f_\theta$ that assigns each sentence $s\in\mathcal{S}_q$ a relevance score $z_s=f_\theta(q,s;\mathcal{S}_q)$, allowing the candidate sentences to be ranked. Top-ranked sentences are then selected and organized as input context for subsequent synthesis generation $y=\operatorname{Gen}_{\phi}(q,\mathcal{S}^{*}_q)$. 

\subsection{Overview}
As illustrated in Figure~\ref{fig:method}, LoG-Reranker adopts a local-to-global reranking paradigm consisting of three components. 
(1) \textit{Local sentence scoring.} Each retrieved passage is first decomposed into sentence-level candidates. Within each query–passage pair, the model encodes the query and sentences, predicts the query intent and sentence roles, and combines their contextual representations with intent–role interactions to compute a fine-grained local relevance score for each sentence.
(2) \textit{Global graph reranking.} The sentence candidates are then organized into a sentence-level graph spanning the entire candidate set. Each node is initialized with its contextual representation, role label, and local relevance score, while intra-passage adjacency edges and cross-passage similarity edges support information propagation. The resulting graph representations are used to compute a global relevance score for each sentence node, producing the final sentence rankings.
(3) \textit{Structured context construction.} Finally, the top-ranked sentence nodes are selected and expanded with their graph neighbors to preserve local context and cross-passage connections, providing a structured input context for downstream generation rather than a flat ranked passage list.
\subsection{Local Sentence Scoring}
For each retrieved passage, the local scoring estimates the fine-grained relevance of its sentences by jointly modeling contextual semantics and the compatibility between query intent and sentence roles (Figure~\ref{fig:method}, left).
\paragraph{Contextual encoding.}
Given a query $q$ and a passage $p_i=\{s_{i,1},\ldots,s_{i,m_i}\}$,we concatenate the query with all sentences and feed them into a shared Transformer encoder:
\begin{equation}
  \mathbf{H}_i
  =
  \operatorname{Encoder}
  \left(
  [q;\,s_{i,1};\ldots;s_{i,m_i}]
  \right),
\end{equation}
where $\mathbf{H}_i\in\mathbb{R}^{L_i\times d}$ denotes the contextualized token representations and $L_i$ is the length of the joint input.
Mean pooling over the corresponding token spans yields the passage-conditioned query representation $\mathbf{h}_{q}^{(i)}$ and contextual sentence representations $\{\mathbf{h}_{i,j}\}_{j=1}^{m_i}$. This joint encoding preserves the local passage context while allowing each sentence to interact with the query.

\paragraph{Intent and role modeling.}
Scientific queries may emphasize different types of information needs, while sentences within scientific passages also serve different functional roles. To explicitly model their compatibility, we predict soft distributions for the query intent and each sentence role:
\begin{equation}
  \boldsymbol{\gamma}_{q}^{(i)}
  =
  \operatorname{softmax}
  \bigl(
  \operatorname{MLP}_{\mathrm{int}}
  (\mathbf{h}_{q}^{(i)})
  \bigr),
\end{equation}
\begin{equation}
  \mathbf{r}_{i,j}
  =
  \operatorname{softmax}
  \bigl(
  \operatorname{MLP}_{\mathrm{role}}
  (\mathbf{h}_{i,j})
  \bigr),
\end{equation}
where $\boldsymbol{\gamma}_{q}^{(i)},\mathbf{r}_{i,j}\in\mathbb{R}^{3}$ denote distributions over background, method, and result.
These soft distributions allow both the query and each sentence to express multiple intent or role characteristics.

Their compatibility is represented through element-wise interaction and projected into a role-aware feature:
\begin{equation}
\mathbf{g}_{i,j}
=
\mathbf{W}_{p}
\left(
\boldsymbol{\gamma}_{q}^{(i)}
\odot
\mathbf{r}_{i,j}
\right),
\end{equation}
where $\odot$ denotes element-wise multiplication and $\mathbf{W}_{p}$ is a learnable projection. For efficient sentence-wise computation, $\boldsymbol{\gamma}_{q}^{(i)}$ is replicated across all sentences in $p_i$, enabling their intent-role interactions and subsequent scoring to be computed in parallel.
\paragraph{Local relevance scoring.}
The local scorer concatenates the contextual query, sentence, and intent-role representations to compute the relevance logit and normalized local score:
\begin{equation}
  \ell_{i,j}^{\mathrm{local}}
  =
  \operatorname{MLP}_{\mathrm{local}}
  \left(
  [\mathbf{h}_{q}^{(i)};
  \mathbf{h}_{i,j};
  \mathbf{g}_{i,j}]
  \right),
\end{equation}
\begin{equation}
  z_{i,j}^{\mathrm{local}}
  =
  \sigma
  \left(
  \ell_{i,j}^{\mathrm{local}}
  \right),
\end{equation}
where $\sigma(\cdot)$ denotes the sigmoid function and the local score $z_{i,j}^{\mathrm{local}}$ jointly reflects contextual query-sentence relevance and intent-role compatibility of $s_{i,j}$. Together with $\mathbf{h}_{i,j}$ and $\mathbf{r}_{i,j}$, they are used to initialize the corresponding node in the global sentence graph.
\subsection{Global Graph Reranking}
Local scoring estimates sentence relevance within individual passages, without considering their relations across the entire candidate set. We therefore construct a query-specific sentence graph $\mathcal{G}_q=(\mathcal{V}_q,\mathcal{E}_q)$ to perform graph-based global reranking (Figure~\ref{fig:method}, middle).
\paragraph{Sentence graph construction.}
Each node $v_{i,j}\in\mathcal{V}_q$ corresponds to a sentence $s_{i,j}$ ($j_{th}$ sentence from $i_{th}$ passage). Its contextual representation $\mathbf{h}_{i,j}$, role distribution $\mathbf{r}_{i,j}$, and local score $z_{i,j}^{\mathrm{local}}$ are separately projected into a shared graph space and fused as node embedding initialization:
\begin{equation}
\mathbf{x}_{i,j}^{(0)}
=
\operatorname{LN}
\left(
\phi_h(\mathbf{h}_{i,j})
+
\phi_r(\mathbf{r}_{i,j})
+
\phi_z(z_{i,j}^{\mathrm{loc}})
\right),
\end{equation}
where $\phi_h$, $\phi_r$, and $\phi_z$ project the features into a shared $d_g$-dimensional space and $LN$ denotes layer normalization.

The edge set consists of intra-passage adjacency edges and cross-passage similarity edges:
\begin{equation}
\mathcal{E}_q
=
\mathcal{E}_{\mathrm{adj}}
\cup
\mathcal{E}_{\mathrm{sim}}.
\end{equation}
For each passage $p_i$, directed adjacency edges connect consecutive sentences in their original order:
\begin{equation}
\mathcal{E}_{\mathrm{adj}}
=
\left\{
(v_{i,j},v_{i,j+1})
\mid
1\leq j<m_i
\right\}.
\end{equation}
Each adjacency edge is typed according to the predicted role label $\hat{r}_{i,j}=\arg\max \mathbf{r}_{i,j}$ of its source
and target nodes, yielding seven relation types: one \textit{same-role} continuity type and six ($A_3^2$) directed \textit{cross-role} transitions among background (\textit{bg}), method (\textit{mt}), and result (\textit{rs}).

For cross-passage connections, we link sentence nodes that share the same predicted role and exhibit similar semantics and local relevance scores:
\begin{equation}
\begin{aligned}
\mathcal{E}_{\mathrm{sim}}
=
\big\{(u,v)\mid\,
& p(u)\neq p(v),\
\hat{r}_u=\hat{r}_v,\\
& c_{uv}\geq\tau_{\mathrm{sim}},\
|z_u^{\mathrm{local}}-z_v^{\mathrm{local}}|\leq\delta,\\
& v\in\operatorname{TopK}_{k}(u)
\big\},
\end{aligned}
\end{equation}
where $p(u)$ denotes the source passage, $c_{uv}=\cos(\mathbf{h}_u,\mathbf{h}_v)$ denotes the cosine similarity between their contextual sentence representations, $\tau_{\mathrm{sim}}$ is the similarity threshold, $\delta$ bounds the local-score difference, and $\operatorname{TopK}_{k}(u)$ retains the $k$ most similar eligible nodes. 

Each retained pair is connected bidirectionally and assigned a shared \textit{similarity} type. Together with the seven intra-passage relation types, this yields edge types set $T$. These constraints restrict cross-passage connections to sentences that are semantically, functionally, and relevance-wise compatible, providing a typed relational structure $\mathcal{G}_q=(\mathcal{V}_q,\mathcal{E}_q,T)$ for subsequent graph encoding.

\paragraph{Relation-aware graph encoding.}
We employ a multi-layer relation-aware graph attention encoder to propagate information over $\mathcal{G}_q$. For an edge $(u,v)$ with relation type $t=rel(u,v)\in T$, the unnormalized and normalized attention weights of head $h$ are computed as:
\begin{equation}
  e_{uv}^{(\ell,h)}
  =
  \operatorname{LeakyReLU}
  \left(
  \mathbf{a}_{h}^{\top}
  [
  \mathbf{W}_{v}\mathbf{x}_{v}^{(\ell)};
  \mathbf{W}_{t}\mathbf{x}_{u}^{(\ell)};
  \mathbf{e}_{t}
  ]
  \right),
\end{equation}
\begin{equation}
  \alpha_{uv}^{(\ell,h)}
  =
  \operatorname{softmax}_{u\in\mathcal{N}(v)}
  \left(
  e_{uv}^{(\ell,h)}
  \right),
\end{equation}
where $\mathbf{W}_{v}$ projects the target node, $\mathbf{W}_{t}$ applies a relation-specific transformation to the source node,  $\mathbf{e}_{t}$ is a learnable embedding of relation type $t$, and $\mathbf{a}_{h}$ denotes the attention parameters of head $h$.

Relation-aware messages are then aggregated across neighbors and attention heads to update each node representation:
\begin{equation}
  \mathbf{m}_{v}^{(\ell,h)}
  =
  \sum_{u\in\mathcal{N}(v)}
  \alpha_{uv}^{(\ell,h)}
  \left[
  \mathbf{W}_{t}\mathbf{x}_{u}^{(\ell)}
  \right]_{h},
\end{equation}
\begin{equation}
  \mathbf{x}_{v}^{(\ell+1)}
  =
  \operatorname{GraphLayer}
  \left(
  \mathbf{x}_{v}^{(\ell)},
  \mathop{\Vert}_{h}\mathbf{m}_{v}^{(\ell,h)}
  \right),
\end{equation}
where each GraphLayer applies an output projection followed by residual normalization and a feed-forward block. After $L$ layers, the resulting representations $\mathbf{X}_{V}^{(L)}=\{\mathbf{x}_{v}^{(L)}\}_{v\in\mathcal{V}_q}$ encode local sentence information and candidate-set relations for global scoring.

\paragraph{Global node scoring.}
The global scorer computes a ranking logit and normalized score from each final node representation:
\begin{equation}
\ell_{i,j}^{\mathrm{global}}
=
\operatorname{MLP}_{\mathrm{global}}
\left(
\mathbf{x}_{i,j}^{(L)}
\right),
\end{equation}
\begin{equation}
z_{i,j}^{\mathrm{global}}
=
\sigma
\left(
\ell_{i,j}^{\mathrm{global}}
\right).
\end{equation}
Since the local score is incorporated into the initial node representation, $z_{i,j}^{\mathrm{global}}$ reflects both query-specific local relevance and the relational structure of the candidate set. The final sentence ranking is obtained by sorting all valid nodes in descending order of $z_{i,j}^{\mathrm{global}}$.

\subsection{Structured Context Construction}
A flat list of isolated sentences may discard the contextual and cross-passage text useful for subsequent generation. Given the global scores, we first select the top-$K$ sentence nodes:
\begin{equation}
\mathcal{V}_{q}^{K}
=
\operatorname{TopK}_{v\in\mathcal{V}_q}
\left(
z_{v}^{\mathrm{global}}
\right).
\end{equation}

For each selected node $v$, we expand it with two directed in-passage neighbors from $\mathcal{E}_{\mathrm{adj}}$ and two most similar cross-passage neighbors from $\mathcal{E}_{\mathrm{sim}}$. 
Each resulting sentence unit $\mathcal{U}(v)$ is identified by $s_{i,j}$, containing its predicted role, sentence text, in-passage adjacency context and cross-passage similarity context (Figure~\ref{fig:method}, right).
The units are ordered by their global scores to form the generation context:
\begin{equation}
\mathcal{C}_q
=
[
\mathcal{U}(v_1),\ldots,\mathcal{U}(v_K)
],
\quad
z_{v_1}^{\mathrm{global}}
\geq\cdots\geq
z_{v_K}^{\mathrm{global}}.
\end{equation}

Thus, the subsequent generator receives fine-grained source information together with its local context and cross-passage connections, rather than an isolated sentence list.

\subsection{Training Objectives}
LoG-Reranker is trained in two stages. We first optimize the local component with sentence-ranking and role supervision, and then freeze it as node initialization to train the global graph reranker.

\paragraph{Local-stage training.}
Let $\mathbf{y}_i=(y_{i,1},\ldots,y_{i,m_i})$ and $\boldsymbol{\ell}_i^{\mathrm{local}} =(\ell_{i,1}^{\mathrm{local}},\ldots,\ell_{i,m_i}^{\mathrm{local}})$ denote the supervision ranks and predicted local logits for sentences in passage $p_i$, respectively, where a smaller rank indicates higher relevance. The local listwise ranking loss is defined as:
\begin{equation}
\widetilde{\mathbf{P}}_i=\operatorname{softmax}(-\mathbf{y}_i/\tau),
\;
\mathbf{P}_i^{\mathrm{local}}
=
\operatorname{softmax}
\left(
\boldsymbol{\ell}_i^{\mathrm{local}}
\right),
\end{equation}
\begin{equation}
  \mathcal{L}_{\mathrm{rank}}^{\mathrm{loc}}
  =
  -\sum_i
  \widetilde{\mathbf{P}}_i^{\top}
  \log\mathbf{P}_i^{\mathrm{local}},
\end{equation}
where $\widetilde{\mathbf{P}}_i$ is the soft target distribution induced by the supervision ranks, $\mathbf{P}_i^{\mathrm{local}}$ is the predicted ranking distribution over sentences in $p_i$, and $\tau$ controls the concentration of the target ranking distribution.

To provide auxiliary supervision for the functional roles of
scientific sentences, we minimize the divergence between the predicted and target role distributions:
\begin{equation}
\mathcal{L}_{\mathrm{role}}
=
\sum_i\sum_{j=1}^{m_i}
D_{\mathrm{KL}}
\left(
\widetilde{\mathbf{r}}_{i,j}
\parallel
\mathbf{r}_{i,j}
\right),
\end{equation}
where $\widetilde{\mathbf{r}}_{i,j}$ denotes the target role distribution of sentence $s_{i,j}$. The overall local-stage objective combines the primary ranking loss with this auxiliary objective:
\begin{equation}
\mathcal{L}_{\mathrm{local}}
=
\mathcal{L}_{\mathrm{rank}}^{\mathrm{loc}}
+
\lambda
\mathcal{L}_{\mathrm{role}},
\end{equation}
where $\lambda$ controls the contribution of auxiliary supervision.

\paragraph{Global-stage training.}
The global reranker is trained with cross-passage preference pairs. For each query, $\mathcal{D}_q$ contains pairs $(u,v)$ from different passages such that $u$ belongs to a higher-ranked passage and has a sentence rank no worse than $v$. The global pairwise ranking loss is:
\begin{equation}
\mathcal{L}_{\mathrm{global}}
=
-\frac{1}{|\mathcal{D}_q|}
\sum_{(u,v)\in\mathcal{D}_q}
\log\sigma
\left(
\ell_u^{\mathrm{global}}
-
\ell_v^{\mathrm{global}}
\right),
\end{equation}
which encourages the preferred node $u$ to receive a higher global ranking logit than $v$.

During this stage, only the graph encoder and global scorer are optimized. Graph construction relies on predicted roles and local scores rather than role labels in supervision data, maintaining consistency between training and inference.
\section{Experiments}
\subsection{Experimental Setup}
\paragraph{Benchmark and Metrics.} We adopt \textit{ScholarQABench}, a benchmark for scientific synthesis, and its task-specific evaluation metrics from OpenScholar \cite{OpenScholar}. It covers three single-paper tasks (PubMedQA, SciFact, and QASA) and four multi-paper tasks (CS, Multi, Bio, and Neuro), with task-specific metrics including accuracy, ROUGE-L, rubric accuracy, citation F1, and organization, relevance, coverage scores based on Prometheus-rubrics \cite{Prometheus}.

Following OpenScholar's retrieval setup, we retrieve 50 passages per query from OS-DataStore using OS-Retriever and add missing gold or curated pseudo-gold evidence for the five labeled tasks (PubMedQA, SciFact, QASA, CS, and Multi), resulting in \textit{ScholarQABench-Rerank}. Bio and Neuro are retained only for generation evaluation because they provide no gold evidence. We evaluate reranking using MAP, Recall, and nDCG. Each reranker's top-10 units are passed to the same OS-8B generator under identical settings \cite{OpenScholar}, and the generated answers are evaluated using the corresponding task-specific metrics in \textit{ScholarQABench}.

\paragraph{Baselines.} We compare our model with nine representative baselines in four families: (1) \textit{Cross-Encoder}, including BGE-Reranker \cite{BGE}, GTE-Reranker \cite{GTE}, and OS-Reranker \cite{OpenScholar}; (2) \textit{LLM-Based}, including RankZephyr \cite{RankZephyr} and CoRank \cite{CoRank}; (3) \textit{Fine-Grained}, including RECOMP \cite{Recomp} and EXIT \cite{EXIT}; and (4) \textit{Graph-Based}, including ChainRAG \cite{ChainRAG} and SentGraph \cite{SentGraph}. Among them, OS-Reranker and CoRank are specialized for scientific literature, and we use Qwen3-Max for baselines requiring an LLM API.

\paragraph{Training Data.}
To construct training supervision, we extract 25,915 single- and multi-paper QA instances from OpenScholar training data \cite{OpenScholar}. We use Qwen3-Max to annotate sentence- and passage-level relevance rankings through batched inference, followed by validation, fragment removal, and rank normalization. A pretrained UniFAR model  \cite{UniFAR} further assigns background, method, and result distributions, resulting in 256,896 passages and 2.14M role-annotated sentences. More details of the training data are provided in the seperate supplementary material.

\paragraph{Implementation Details.}
During training, \textit{gte-reranker-modernbert-base} serves as the base encoder of the local reranker, with a maximum input length of 1,024.
The local reranker is trained for five epochs using a learning rate of $2\times10^{-5}$, a micro-batch size of 2, and a role-loss weight of 0.5; its best checkpoint is frozen for global training. The global reranker employs a two-layer, four-head relation-aware GAT with a hidden dimension of 256. Cross-passage edges retain the top-8 same-role neighbors with $\tau_{\mathrm{sim}}=0.75$ and $\delta=0.25$. The global reranker is trained for five epochs using a learning rate of $3\times10^{-5}$ and a micro-batch size of 1. Both stages use AdamW, eight-step gradient accumulation, 0.01 weight decay, 0.1 dropout, 6\% warmup, and a random seed of 13. More implementation details are provided in the supplementary material.

\begin{table*}[t!]
  \small
  \centering
  \setlength{\tabcolsep}{3.8pt}
  \renewcommand{\arraystretch}{0.93}
  \begin{tabular}{@{}cccccccccccccccc@{}}
    \toprule
    \multirow{2}{*}{Model   / Metric} & \multicolumn{3}{c}{PubMedQA} & \multicolumn{3}{c}{SciFact} & \multicolumn{3}{c}{QASA} & \multicolumn{3}{c}{Scholar-CS} & \multicolumn{3}{c}{Scholar-Multi} \\ 
    \cmidrule(l){2-4} \cmidrule(l){5-7} \cmidrule(l){8-10} \cmidrule(l){11-13} \cmidrule(l){14-16}
     & Recall & MAP & nDCG & Recall & MAP & nDCG & Recall & MAP & nDCG & Recall & MAP & nDCG & Recall & MAP & nDCG \\
     \midrule
    BGE-Reranker & 89.26 & 73.24 & 79.19 & 74.61 & 60.84 & 67.06 & 75.43 & 48.57 & 54.81 & 95.00 & 88.87 & 90.95 & 39.37 & 24.13 & 34.55 \\
    GTE-Reranker & 90.32 & 75.04 & 80.78 & 75.96 & 61.42 & 67.32 & 78.73 & 53.56 & 59.04 & 95.00 & 88.74 & 90.88 & 44.50 & 28.33 & 39.48 \\
    OS-Reranker & 91.92 & 81.34 & 84.88 & 80.18 & 63.38 & 71.71 & 83.14 & 60.69 & 65.04 & \underline{95.50} & \underline{89.42} & \underline{91.86} & 45.31 & 28.73 & 39.80 \\
    RankZephyr & 61.94 & 48.49 & 55.30 & 56.92 & 48.14 & 52.98 & 66.74 & 42.08 & 50.45 & 85.50 & 75.71 & 81.97 & 30.73 & 21.09 & 28.97 \\
    CoRank & \underline{92.81} & \underline{83.08} & \underline{86.95} & \underline{85.60} & \underline{66.49} & \underline{75.08} & \underline{85.67} & 61.72 & \textbf{67.10} & \textbf{95.75} & 89.25 & 91.46 & 43.98 & 27.75 & 38.14 \\
    RECOMP & 80.98 & 57.85 & 64.26 & 63.61 & 52.92 & 60.88 & 77.46 & 50.94 & 56.63 & 93.75 & 87.86 & 89.23 & 31.50 & 21.61 & 29.90 \\
    EXIT & 83.89 & 59.26 & 67.84 & 70.98 & 55.99 & 62.27 & 69.68 & 43.38 & 51.94 & 90.50 & 83.19 & 85.66 & 41.40 & 25.28 & 36.41 \\
    ChainRAG & 72.99 & 50.42 & 55.20 & 63.32 & 51.95 & 58.55 & 70.78 & 43.41 & 51.89 & 94.75 & 87.79 & 90.21 & 33.42 & 22.86 & 31.20 \\
    SentGraph & 85.17 & 64.35 & 77.15 & 83.96 & 64.50 & 74.31 & 85.08 & \underline{62.23} & \underline{67.08} & 95.25 & 89.14 & 91.32 & \underline{45.89} & \underline{29.15} & \underline{40.69} \\
    \cmidrule(l){1-16} 
    LoG-Reranker & \textbf{93.63} & \textbf{84.73} & \textbf{88.83} & \textbf{87.47} & \textbf{67.85} & \textbf{76.14} & \textbf{86.42} & \textbf{63.98} & 66.81 & \textbf{95.75} & \textbf{89.76} & \textbf{92.04} & \textbf{46.87} & \textbf{30.46} & \textbf{41.29} \\
    \bottomrule
    \end{tabular}
    \caption{Reranking performance on ScholarQABench-Rerank. Recall, MAP, and nDCG are reported over the top-10 reranked units (@10). Best and second-best results are shown in bold and underlined, respectively.}
    \label{tab:reranking}
\end{table*}

\begin{table*}[t!]
  \small
  \centering
  \setlength{\tabcolsep}{3.5pt}
  \renewcommand{\arraystretch}{0.93}
  \begin{tabular}{@{}ccccccccccccccc@{}}
  \toprule
  \multirow{3}{*}{\begin{tabular}[c]{@{}c@{}}Task\\ \\ Model   / Metric\end{tabular}} & \multicolumn{6}{c}{Single-Paper} & \multicolumn{8}{c}{Multi-Paper} \\ \cmidrule(l){2-7} \cmidrule(l){8-15}
   & \multicolumn{2}{c}{PubMedQA} & \multicolumn{2}{c}{SciFact} & \multicolumn{2}{c}{QASA} & \multicolumn{2}{c}{Scholar-CS} & \multicolumn{2}{c}{Scholar-Multi} & \multicolumn{2}{c}{Scholar-Bio} & \multicolumn{2}{c}{Scholar-Neuro} \\ \cmidrule(l){2-3} \cmidrule(l){4-5} \cmidrule(l){6-7} \cmidrule(l){8-9} \cmidrule(l){10-11} \cmidrule(l){12-13}
   \cmidrule(l){14-15}
   & Acc & Cite & Acc & Cite & R-L & Cite & Rub-Acc & Cite & Rub-Score & Cite & Rub-Score & Cite & Rub-Score & Cite \\ \midrule
  BGE-Reranker & 82.44 & 76.12 & 81.73 & 50.24 & 19.16 & 45.62 & 60.86 & 36.23 & 3.910 & \underline{52.00} & 4.131 & 51.81 & 4.058 & 52.33 \\
  GTE-Reranker & 82.44 & 75.21 & \underline{82.69} & 52.88 & 19.29 & 46.11 & 59.56 & 31.02 & \underline{3.975} & 49.97 & 4.140 & \underline{53.91} & 4.082 & 52.70 \\
  OS-Reranker & \underline{83.27} & \underline{76.66} & 76.92 & \underline{53.95} & 18.79 & 45.93 & \underline{61.72} & 37.08 & 3.923 & 51.71 & \underline{4.163} & 52.11 & 4.088 & 52.17 \\
  RankZephyr & 67.14 & 54.39 & 73.56 & 41.83 & 17.24 & 40.97 & 58.70 & 32.68 & 3.871 & 47.24 & 4.125 & 48.13 & 4.046 & 50.56 \\
  CoRank & 82.56 & 74.28 & 80.25 & 53.17 & 19.02 & \underline{46.24} & 61.15 & 35.89 & 3.964 & 50.95 & 4.155 & 51.77 & \underline{4.097} & \underline{52.85} \\
  RECOMP & 69.99 & 64.56 & 74.52 & 38.46 & 18.46 & 33.26 & 55.39 & 33.81 & 3.735 & 41.36 & 4.055 & 38.39 & 3.983 & 40.19 \\
  EXIT & 69.04 & 65.56 & 81.62 & 50.12 & \underline{19.48} & 40.41 & 54.80 & 37.51 & 3.682 & 37.16 & 4.092 & 40.29 & 4.043 & 41.75 \\
  ChainRAG & 71.41 & 67.97 & 78.37 & 47.91 & 18.63 & 41.56 & 59.47 & \underline{38.76} & 3.753 & 48.98 & 4.117 & 47.12 & 4.037 & 46.67 \\
  SentGraph & 72.95 & 70.83 & 79.81 & 48.29 & 18.35 & 42.18 & 61.32 & 38.14 & 3.797 & 49.23 & 4.124 & 50.65 & 4.033 & 49.84 \\
  \cmidrule(l){1-15}
  LoG-Reranker & \textbf{84.67} & \textbf{77.92} & \textbf{83.17} & \textbf{55.10} & \textbf{20.38} & \textbf{47.76} & \textbf{62.44} & \textbf{39.42} & \textbf{4.006} & \textbf{52.74} & \textbf{4.240} & \textbf{54.87} & \textbf{4.152} & \textbf{53.80} \\
  \bottomrule
  \end{tabular}
  \caption{Generation performance on ScholarQABench using the top-10 reranked units. Task-specific metrics include accuracy (Acc) for PubMedQA and SciFact, ROUGE-L (R-L) for QASA, rubric accuracy (Rub-Acc) for Scholar-CS, and the aggregated Prometheus rubric score (Rub-Score) over organization,relevance, and coverage for the remaining tasks; Cite denotes citation F1. Best and second-best results are shown in bold and underlined, respectively.}
  \label{tab:generation}
\end{table*}

\subsection{Main Results}
Tables~\ref{tab:reranking} and~\ref{tab:generation} compare LoG-Reranker with representative baselines from two complementary perspectives, reranking and generation. We summarize two key
observations:

\textbf{(1) Fine-grained relational modeling consistently improves reranking.}
As shown in Table~\ref{tab:reranking}, LoG-Reranker achieves the best result in 14 of the 15 task-metric comparisons, including one tie. The strongest competing approach varies substantially across datasets: CoRank is highly competitive on PubMedQA and SciFact, OS-Reranker performs strongly on Scholar-CS, and SentGraph provides the strongest baseline results on Scholar-Multi. This pattern shows that no single competing paradigm performs uniformly well across the evaluated datasets. In contrast, LoG-Reranker remains consistently effective under these differing settings. For example, it improves MAP over the strongest baseline by 1.36 points on SciFact and 1.31 points on Scholar-Multi. The gains span Recall, MAP, and nDCG, indicating that LoG-Reranker both retains more relevant information within the top-10 cutoff and ranks it more accurately. Although its nDCG on QASA is slightly below the best result, it still obtains the highest Recall and MAP on that dataset.
Overall, these results show that combining sentence-level local relevance with candidate-set relational structure provides robust improvements over competitive rerankers.

\textbf{(2) Improved reranking leads to more effective scientific synthesis generation.}
Table~\ref{tab:generation} provides a stronger evaluation of whether the ranking improvements are useful beyond the reranking objective. LoG-Reranker obtains the best task-specific generation score and citation F1 on all seven benchmarks, covering every reported comparison. Since all methods use the same OS-8B generator and top-10 input setting, the improvements reflect the quality of the input contexts produced before generation rather than differences in generator capacity. Notably, the gains occur simultaneously in task-specific metrics and citation F1. On QASA, LoG-Reranker improves ROUGE-L by 0.90 points and citation F1 by 1.52 points over the strongest corresponding baselines, while on Scholar-CS it improves rubric accuracy and citation F1 by 0.72 and 0.66 points, respectively. In terms of the quality of generated answer, LoG-Reranker also achieves the highest Rub-Score on Scholar-Multi, Scholar-Bio, and Scholar-Neuro, reflecting stronger organization, relevance, and coverage in scientific synthesis.
These results demonstrate that the benefits extend beyond ranking metrics, as fine-grained information prioritization and relation-aware context organization provide the generator with source input that better supports answer quality, coverage, and citation grounding.

\begin{figure*}[t!]
  \centering
  \includegraphics[width=0.98\textwidth]{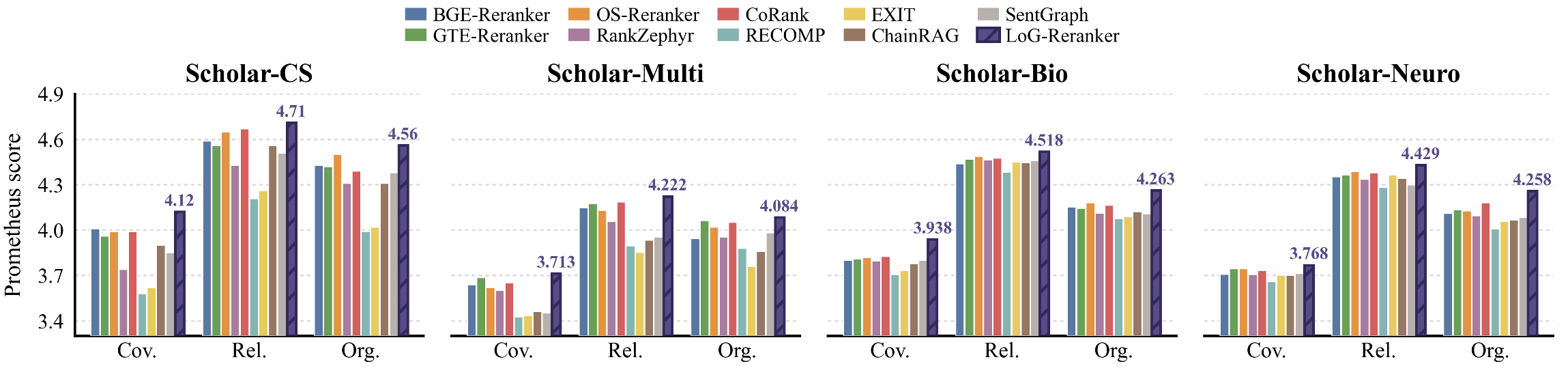}
  \caption{Comparison of Prometheus rubric scores across four scientific synthesis tasks, where Cov., Rel., and Org. denote coverage, relevance, and organization, respectively.}
  \label{fig:details}
\end{figure*}

\begin{table*}[!t]
  \small
  \centering
  \setlength{\tabcolsep}{8pt}
  \renewcommand{\arraystretch}{0.96}
  \begin{tabular}{@{}ccccccccccccc@{}}
  \toprule
  \multirow{2}{*}{Model   / Metric} & \multicolumn{3}{c}{Scholar-CS} & \multicolumn{3}{c}{Scholar-Multi} & \multicolumn{3}{c}{Scholar-Bio} & \multicolumn{3}{c}{Scholar-Neuro} \\
  \cmidrule(l){2-4} \cmidrule(l){5-7} \cmidrule(l){8-10} \cmidrule(l){11-13}
   & Cov. & Rel. & Org. & Cov. & Rel. & Org. & Cov. & Rel. & Org. & Cov. & Rel. & Org. \\
  \midrule
  LoG-Reranker & \textbf{4.12} & \textbf{4.71} & \textbf{4.56} & \textbf{3.713} & \textbf{4.222} & \textbf{4.084} & \textbf{3.938} & \textbf{4.518} & \textbf{4.263} & \textbf{3.768} & \textbf{4.429} & \textbf{4.258} \\
  \textit{w/o Local} & 3.38 & 4.15 & 3.90 & 3.365 & 3.793 & 3.803 & 3.687 & 4.387 & 4.085 & 3.667 & 4.268 & 4.082 \\
  \textit{w/o Global} & 3.46 & 4.22 & 3.95 & 3.407 & 3.917 & 3.806 & 3.717 & 4.461 & 4.022 & 3.689 & 4.362 & 4.013 \\
  \textit{w/o Structured} & 3.70 & 4.38 & 4.14 & 3.602 & 4.176 & 3.898 & 3.782 & 4.507 & 4.178 & 3.719 & 4.397 & 4.137 \\
  \bottomrule
  \end{tabular}
  \caption{Ablation results on multi-paper synthesis tasks. Cov., Rel., and Org. denote coverage, relevance, and organization scores evaluated by Prometheus rubrics, respectively.}
  \label{tab:ablation}
\end{table*}
\subsection{Analysis and Ablation Study}
To better understand the generation improvements of LoG-Reranker, we conduct a
fine-grained quality analysis (Figure~\ref{fig:details}) and a component-wise
ablation study (Table~\ref{tab:ablation}).

\textbf{Improvements span all synthesis dimensions.}
As shown in Figure~\ref{fig:details}, LoG-Reranker achieves the highest score in all 12 task-dimension comparisons, indicating that its generation gains are consistent across tasks and rubric dimensions rather than concentrated in a particular aspect of synthesis quality. Relative to the strongest baseline, the most pronounced margins appear in coverage on Scholar-CS and Scholar-Bio and in organization on Scholar-Bio and Scholar-Neuro. This pattern is consistent with the local-to-global design: fine-grained sentence selection helps retain useful information, while relation-aware reranking and context construction facilitate its integration across papers. Importantly, relevance
also improves on every task, showing that the gains in coverage and organization are achieved without diluting the focus of the generated answers.

\textbf{Local and global modeling provide complementary benefits.}
Table~\ref{tab:ablation} shows that removing either reranking stage reduces all three rubric scores across every task. \textit{w/o Local} removes role-aware sentence modeling and local relevance scoring, replacing sentence nodes with passage nodes connected only by semantic similarity. This variant produces
the largest overall degradation, with coverage decreasing by 0.360 points on average, confirming that passage-level representations cannot adequately expose the fine-grained information required for synthesis. In contrast, \textit{w/o Global} retains the local sentence scorer but removes the graph encoder and global scorer, ranking sentences solely by their local scores. Despite preserving the same sentence candidates, it still causes substantial declines, with organization decreasing by 0.344 points on average compared with 0.230 for relevance. This difference indicates that independently relevant sentences do not necessarily form a coherent candidate set and global graph modeling complements local relevance by accounting for intra-passage context and cross-passage relations.

\textbf{Structured context contributes beyond sentence ranking.}
The \textit{w/o Structured} variant in Table~\ref{tab:ablation} preserves the complete local-to-global reranking process but provides the generator with a flat list of top-ranked sentences without adjacency or similarity neighbor expansion. Because the underlying rankings remain unchanged, its performance difference isolates the contribution of relation-aware context construction from that of sentence selection. Nevertheless, \textit{w/o Structured} underperforms the complete model in all 12 comparisons. Averaged across the four tasks, organization and coverage decrease by 0.203 and 0.184 points, respectively, compared with 0.105 for relevance. This pattern reflects the
different functions of graph neighbors: adjacency neighbors preserve local continuity, while cross-passage neighbors supplement related information from other sources. Retaining these relations therefore provides a more coherent and well-covered generation context than a flat sentence list.

\section{Conclusion}
In this work, we present LoG-Reranker, a local-to-global sentence-level reranking framework for scientific synthesis, integrating fine-grained semantic relevance with relational structure. By jointly modeling local query-sentence interactions and global sentence-level graph relations, it provides the generator with a more informative and structured context, enabling more grounded and comprehensive synthesis. Experimental results on scientific synthesis benchmarks show consistent improvements over competitive reranking baselines, while ablation studies further validate the complementary contributions of the proposed components.

\bibliography{refs}

@article{OpenScholar,
  title   = {Synthesizing scientific literature with retrieval-augmented language models},
  author  = {Asai, Akari and He, Jacqueline and Shao, Rulin and others},
  journal = {Nature},
  volume  = {650},
  pages   = {857--863},
  year    = {2026},
  doi     = {10.1038/s41586-025-10072-4},
  url     = {https://doi.org/10.1038/s41586-025-10072-4}
}

@inproceedings{SciRAG,
    title = "{S}ci{RAG}: Adaptive, Citation-Aware, and Outline-Guided Retrieval and Synthesis for Scientific Literature",
    author = "Ding, Hang  and
      Zhao, Yilun  and
      Hu, Tiansheng  and
      Wang, Zihang  and
      Patwardhan, Manasi  and
      Cohan, Arman",
    booktitle = "Proceedings of the 19th Conference of the {E}uropean Chapter of the {A}ssociation for {C}omputational {L}inguistics (Volume 1: Long Papers)",
    month = mar,
    year = "2026",
    address = "Rabat, Morocco",
    publisher = "Association for Computational Linguistics",
    url = "https://aclanthology.org/2026.eacl-long.303/",
    doi = "10.18653/v1/2026.eacl-long.303",
    pages = "6440--6460",
    ISBN = "979-8-89176-380-7"
}

@article{LitLLMs,
  title     = {LitLLMs, LLMs for Literature Review: Are We There yet?},
  author    = {Agarwal, Shubham and Sahu, Gaurav and Puri, Abhay and Laradji, Issam H. and Dvijotham, Krishnamurthy Dj and Stanley, Jason and Charlin, Laurent and Pal, Christopher},
  journal   = {Transactions on Machine Learning Research},
  year      = {2025},
  url       = {https://openreview.net/pdf/8c3bf30be5aecf9b0413b44043f401df8b2d1a3e.pdf}
}

@inproceedings{GTE,
  title={mGTE: Generalized Long-Context Text Representation and Reranking Models for Multilingual Text Retrieval},
  author={Zhang, Xin and Zhang, Yanzhao and Long, Dingkun and Xie, Wen and Dai, Ziqi and Tang, Jialong and Lin, Huan and Yang, Baosong and Xie, Pengjun and Huang, Fei and others},
  booktitle={Proceedings of the 2024 Conference on Empirical Methods in Natural Language Processing: Industry Track},
  pages={1393--1412},
  year={2024}
}

@misc{BGE,
      title={BGE M3-Embedding: Multi-Lingual, Multi-Functionality, Multi-Granularity Text Embeddings Through Self-Knowledge Distillation}, 
      author={Jianlv Chen and Shitao Xiao and Peitian Zhang and Kun Luo and Defu Lian and Zheng Liu},
      year={2024},
      eprint={2402.03216},
      archivePrefix={arXiv},
      primaryClass={cs.CL}
}

@inproceedings{ReClaim,
    title = "Ground Every Sentence: Improving Retrieval-Augmented {LLM}s with Interleaved Reference-Claim Generation",
    author = "Xia, Sirui  and
      Wang, Xintao  and
      Liang, Jiaqing  and
      Zhang, Yifei  and
      Zhou, Weikang  and
      Deng, Jiaji  and
      Yu, Fei  and
      Xiao, Yanghua",
    booktitle = "Findings of the Association for Computational Linguistics: NAACL 2025",
    month = apr,
    year = "2025",
    address = "Albuquerque, New Mexico",
    publisher = "Association for Computational Linguistics",
    url = "https://aclanthology.org/2025.findings-naacl.55/",
    doi = "10.18653/v1/2025.findings-naacl.55",
    pages = "969--988",
    ISBN = "979-8-89176-195-7"
}

@inproceedings{LongCite,
    title = "{L}ong{C}ite: Enabling {LLM}s to Generate Fine-grained Citations in Long-Context {QA}",
    author = "Zhang, Jiajie  and
      Bai, Yushi  and
      Lv, Xin  and
      Gu, Wanjun  and
      Liu, Danqing  and
      Zou, Minhao  and
      Cao, Shulin  and
      Hou, Lei  and
      Dong, Yuxiao  and
      Feng, Ling  and
      Li, Juanzi",
    booktitle = "Findings of the Association for Computational Linguistics: ACL 2025",
    month = jul,
    year = "2025",
    address = "Vienna, Austria",
    publisher = "Association for Computational Linguistics",
    url = "https://aclanthology.org/2025.findings-acl.264/",
    doi = "10.18653/v1/2025.findings-acl.264",
    pages = "5098--5122",
    ISBN = "979-8-89176-256-5"
}

@article{Lost,
    title = "Lost in the Middle: How Language Models Use Long Contexts",
    author = "Liu, Nelson F.  and
      Lin, Kevin  and
      Hewitt, John  and
      Paranjape, Ashwin  and
      Bevilacqua, Michele  and
      Petroni, Fabio  and
      Liang, Percy",
    journal = "Transactions of the Association for Computational Linguistics",
    volume = "12",
    year = "2024",
    address = "Cambridge, MA",
    publisher = "MIT Press",
    url = "https://aclanthology.org/2024.tacl-1.9/",
    doi = "10.1162/tacl_a_00638",
    pages = "157--173"
}

@inproceedings{LongFormQAEval,
    title = "A Critical Evaluation of Evaluations for Long-form Question Answering",
    author = "Xu, Fangyuan  and
      Song, Yixiao  and
      Iyyer, Mohit  and
      Choi, Eunsol",
    booktitle = "Proceedings of the 61st Annual Meeting of the Association for Computational Linguistics (Volume 1: Long Papers)",
    month = jul,
    year = "2023",
    address = "Toronto, Canada",
    publisher = "Association for Computational Linguistics",
    url = "https://aclanthology.org/2023.acl-long.181/",
    doi = "10.18653/v1/2023.acl-long.181",
    pages = "3225--3245"
}

@inproceedings{ICAT,
    title = "Beyond Factual Accuracy: Evaluating Coverage of Diverse Factual Information in Long-form Text Generation",
    author = "Samarinas, Chris  and
      Krubner, Alexander  and
      Salemi, Alireza  and
      Kim, Youngwoo  and
      Zamani, Hamed",
    booktitle = "Findings of the Association for Computational Linguistics: ACL 2025",
    month = jul,
    year = "2025",
    address = "Vienna, Austria",
    publisher = "Association for Computational Linguistics",
    url = "https://aclanthology.org/2025.findings-acl.693/",
    doi = "10.18653/v1/2025.findings-acl.693",
    pages = "13468--13482",
    ISBN = "979-8-89176-256-5"
}

@misc{CrossEncoder,
      title={Passage Re-ranking with BERT}, 
      author={Rodrigo Nogueira and Kyunghyun Cho},
      year={2020},
      eprint={1901.04085},
      archivePrefix={arXiv},
      primaryClass={cs.IR},
      url={https://arxiv.org/abs/1901.04085}, 
}

@misc{SequenceToSequence,
      title={Document Ranking with a Pretrained Sequence-to-Sequence Model}, 
      author={Rodrigo Nogueira and Zhiying Jiang and Jimmy Lin},
      year={2020},
      eprint={2003.06713},
      archivePrefix={arXiv},
      primaryClass={cs.IR},
      url={https://arxiv.org/abs/2003.06713}, 
}

@inproceedings{RankT5,
author = {Zhuang, Honglei and Qin, Zhen and Jagerman, Rolf and Hui, Kai and Ma, Ji and Lu, Jing and Ni, Jianmo and Wang, Xuanhui and Bendersky, Michael},
title = {RankT5: Fine-Tuning T5 for Text Ranking with Ranking Losses},
year = {2023},
isbn = {9781450394086},
publisher = {Association for Computing Machinery},
address = {New York, NY, USA},
url = {https://doi.org/10.1145/3539618.3592047},
doi = {10.1145/3539618.3592047},
booktitle = {Proceedings of the 46th International ACM SIGIR Conference on Research and Development in Information Retrieval},
pages = {2308–2313},
numpages = {6},
location = {Taipei, Taiwan},
}

@inproceedings{ListT5,
    title = "{L}ist{T}5: Listwise Reranking with Fusion-in-Decoder Improves Zero-shot Retrieval",
    author = "Yoon, Soyoung  and
      Choi, Eunbi  and
      Kim, Jiyeon  and
      Yun, Hyeongu  and
      Kim, Yireun  and
      Hwang, Seung-won",
    booktitle = "Proceedings of the 62nd Annual Meeting of the Association for Computational Linguistics (Volume 1: Long Papers)",
    month = aug,
    year = "2024",
    address = "Bangkok, Thailand",
    publisher = "Association for Computational Linguistics",
    url = "https://aclanthology.org/2024.acl-long.125/",
    doi = "10.18653/v1/2024.acl-long.125",
    pages = "2287--2308",
}

@misc{ZeroShot,
      title={Zero-Shot Listwise Document Reranking with a Large Language Model}, 
      author={Xueguang Ma and Xinyu Zhang and Ronak Pradeep and Jimmy Lin},
      year={2023},
      eprint={2305.02156},
      archivePrefix={arXiv},
      primaryClass={cs.IR},
      url={https://arxiv.org/abs/2305.02156}, 
}

@inproceedings{RankGPT,
    title = "Is {C}hat{GPT} Good at Search? Investigating Large Language Models as Re-Ranking Agents",
    author = "Sun, Weiwei  and
      Yan, Lingyong  and
      Ma, Xinyu  and
      Wang, Shuaiqiang  and
      Ren, Pengjie  and
      Chen, Zhumin  and
      Yin, Dawei  and
      Ren, Zhaochun",
    booktitle = "Proceedings of the 2023 Conference on Empirical Methods in Natural Language Processing",
    month = dec,
    year = "2023",
    address = "Singapore",
    publisher = "Association for Computational Linguistics",
    url = "https://aclanthology.org/2023.emnlp-main.923/",
    doi = "10.18653/v1/2023.emnlp-main.923",
    pages = "14918--14937"
}

@misc{RankZephyr,
      title={RankZephyr: Effective and Robust Zero-Shot Listwise Reranking is a Breeze!}, 
      author={Ronak Pradeep and Sahel Sharifymoghaddam and Jimmy Lin},
      year={2023},
      eprint={2312.02724},
      archivePrefix={arXiv},
      primaryClass={cs.IR},
      url={https://arxiv.org/abs/2312.02724}, 
}

@misc{DeepEra,
      title={DeepEra: A Deep Evidence Reranking Agent for Scientific Retrieval-Augmented Generated Question Answering}, 
      author={Haotian Chen and Qingqing Long and Siyu Pu and Xiao Luo and Wei Ju and Meng Xiao and Yuanchun Zhou and Jianghua Zhao and Xuezhi Wang},
      year={2026},
      eprint={2601.16478},
      archivePrefix={arXiv},
      primaryClass={cs.CL},
      url={https://arxiv.org/abs/2601.16478}, 
}

@inproceedings{RECOMP,
  author       = {Fangyuan Xu and
                  Weijia Shi and
                  Eunsol Choi},
  title        = {{RECOMP:} Improving Retrieval-Augmented LMs with Context Compression
                  and Selective Augmentation},
  booktitle    = {The Twelfth International Conference on Learning Representations,
                  {ICLR} 2024, Vienna, Austria, May 7-11, 2024},
  year         = {2024},
  url          = {https://openreview.net/forum?id=mlJLVigNHp},
}

@inproceedings{DSLR,
    title = "{DSLR}: Document Refinement with Sentence-Level Re-ranking and Reconstruction to Enhance Retrieval-Augmented Generation",
    author = "Hwang, Taeho  and
      Jeong, Soyeong  and
      Cho, Sukmin  and
      Han, SeungYoon  and
      Park, Jong",
    booktitle = "Proceedings of the 3rd Workshop on Knowledge Augmented Methods for NLP",
    month = aug,
    year = "2024",
    address = "Bangkok, Thailand",
    publisher = "Association for Computational Linguistics",
    url = "https://aclanthology.org/2024.knowledgenlp-1.6/",
    doi = "10.18653/v1/2024.knowledgenlp-1.6",
    pages = "73--92"
}

@inproceedings{EXIT,
    title = "{EXIT}: Context-Aware Extractive Compression for Enhancing Retrieval-Augmented Generation",
    author = "Hwang, Taeho  and
      Cho, Sukmin  and
      Jeong, Soyeong  and
      Song, Hoyun  and
      Han, SeungYoon  and
      Park, Jong C.",
    booktitle = "Findings of the Association for Computational Linguistics: ACL 2025",
    month = jul,
    year = "2025",
    address = "Vienna, Austria",
    publisher = "Association for Computational Linguistics",
    url = "https://aclanthology.org/2025.findings-acl.253/",
    doi = "10.18653/v1/2025.findings-acl.253",
    pages = "4895--4924",
    ISBN = "979-8-89176-256-5"
}

@inproceedings{ECoRAG,
    title = "{EC}o{RAG}: Evidentiality-guided Compression for Long Context {RAG}",
    author = "Jeong, Yeonseok  and
      Kim, Jinsu  and
      Lee, Dohyeon  and
      Hwang, Seung-won",
    booktitle = "Findings of the Association for Computational Linguistics: ACL 2025",
    month = jul,
    year = "2025",
    address = "Vienna, Austria",
    publisher = "Association for Computational Linguistics",
    url = "https://aclanthology.org/2025.findings-acl.1365/",
    doi = "10.18653/v1/2025.findings-acl.1365",
    pages = "26607--26628",
    ISBN = "979-8-89176-256-5"
}

@inproceedings{Fine-grained,
    title = "Fine-grained Knowledge Enhancement for Retrieval-Augmented Generation",
    author = "Han, Jingxuan  and
      Mao, Zhendong  and
      Liu, Yi  and
      Che, Yexuan  and
      Fu, Zheren  and
      Wang, Quan",
    booktitle = "Findings of the Association for Computational Linguistics: ACL 2025",
    month = jul,
    year = "2025",
    address = "Vienna, Austria",
    publisher = "Association for Computational Linguistics",
    url = "https://aclanthology.org/2025.findings-acl.522/",
    doi = "10.18653/v1/2025.findings-acl.522",
    pages = "10031--10044",
    ISBN = "979-8-89176-256-5"
}

@misc{CoRank,
      title={CoRank: LLM-Based Compact Reranking with Document Features for Scientific Retrieval}, 
      author={Runchu Tian and Xueqiang Xu and Bowen Jin and SeongKu Kang and Jiawei Han},
      year={2025},
      eprint={2505.13757},
      archivePrefix={arXiv},
      primaryClass={cs.IR},
      url={https://arxiv.org/abs/2505.13757}, 
}

@misc{dont,
      title={Don't Forget to Connect! Improving RAG with Graph-based Reranking}, 
      author={Jialin Dong and Bahare Fatemi and Bryan Perozzi and Lin F. Yang and Anton Tsitsulin},
      year={2024},
      eprint={2405.18414},
      archivePrefix={arXiv},
      primaryClass={cs.CL},
      url={https://arxiv.org/abs/2405.18414}, 
}

@inproceedings{GAR,
author = {MacAvaney, Sean and Tonellotto, Nicola and Macdonald, Craig},
title = {Adaptive Re-Ranking with a Corpus Graph},
year = {2022},
isbn = {9781450392365},
publisher = {Association for Computing Machinery},
address = {New York, NY, USA},
url = {https://doi.org/10.1145/3511808.3557231},
doi = {10.1145/3511808.3557231},
booktitle = {Proceedings of the 31st ACM International Conference on Information and Knowledge Management},
pages = {1491–1500},
numpages = {10},
location = {Atlanta, GA, USA},
series = {CIKM '22}
}

@misc{GNRR,
      title={Graph Neural Re-Ranking via Corpus Graph}, 
      author={Andrea Giuseppe Di Francesco and Christian Giannetti and Nicola Tonellotto and Fabrizio Silvestri},
      year={2024},
      eprint={2406.11720},
      archivePrefix={arXiv},
      primaryClass={cs.IR},
      url={https://arxiv.org/abs/2406.11720}, 
}

@inproceedings{KERM,
author = {Dong, Qian and Liu, Yiding and Cheng, Suqi and Wang, Shuaiqiang and Cheng, Zhicong and Niu, Shuzi and Yin, Dawei},
title = {Incorporating Explicit Knowledge in Pre-trained Language Models for Passage Re-ranking},
year = {2022},
isbn = {9781450387323},
publisher = {Association for Computing Machinery},
address = {New York, NY, USA},
url = {https://doi.org/10.1145/3477495.3531997},
doi = {10.1145/3477495.3531997},
booktitle = {Proceedings of the 45th International ACM SIGIR Conference on Research and Development in Information Retrieval},
pages = {1490–1501},
numpages = {12},
location = {Madrid, Spain},
series = {SIGIR '22}
}

@misc{GraphER,
    title={GraphER: An Efficient Graph-Based Enrichment and Reranking Method for Retrieval-Augmented Generation}, 
    author={Ruizhong Miao and Yuying Wang and Rongguang Wang and Chenyang Li and Tao Sheng and Sujith Ravi and Dan Roth},
    year={2026},
    eprint={2603.24925},
    archivePrefix={arXiv},
    primaryClass={cs.LG},
    url={https://arxiv.org/abs/2603.24925}, 
}

@inproceedings{Joint,
    title = "Question-Answer Sentence Graph for Joint Modeling Answer Selection",
    author = "Iyer, Roshni  and
      Vu, Thuy  and
      Moschitti, Alessandro  and
      Sun, Yizhou",
    booktitle = "Proceedings of the 17th Conference of the European Chapter of the Association for Computational Linguistics",
    month = may,
    year = "2023",
    address = "Dubrovnik, Croatia",
    publisher = "Association for Computational Linguistics",
    url = "https://aclanthology.org/2023.eacl-main.68/",
    doi = "10.18653/v1/2023.eacl-main.68",
    pages = "968--979"
}

@inproceedings{ChainRAG,
    title = "Mitigating Lost-in-Retrieval Problems in Retrieval Augmented Multi-Hop Question Answering",
    author = "Zhu, Rongzhi  and
      Liu, Xiangyu  and
      Sun, Zequn  and
      Wang, Yiwei  and
      Hu, Wei",
    booktitle = "Proceedings of the 63rd Annual Meeting of the Association for Computational Linguistics (Volume 1: Long Papers)",
    month = jul,
    year = "2025",
    address = "Vienna, Austria",
    publisher = "Association for Computational Linguistics",
    url = "https://aclanthology.org/2025.acl-long.1089/",
    doi = "10.18653/v1/2025.acl-long.1089",
    pages = "22362--22375",
    ISBN = "979-8-89176-251-0"
}

@misc{SentGraph,
      title={SentGraph: Hierarchical Sentence Graph for Multi-hop Retrieval-Augmented Question Answering}, 
      author={Junli Liang and Pengfei Zhou and Wangqiu Zhou and Wenjie Qing and Qi Zhao and Ziwen Wang and Qi Song and Xiangyang Li},
      year={2026},
      eprint={2601.03014},
      archivePrefix={arXiv},
      primaryClass={cs.CL},
      url={https://arxiv.org/abs/2601.03014}, 
}

@misc{UniFAR,
      title={UniFAR: A Unified Facet-Aware Retrieval Framework for Scientific Documents}, 
      author={Zheng Dou and Zhao Zhang and Deqing Wang and Yikun Ban and Fuzhen Zhuang},
      year={2026},
      eprint={2602.23766},
      archivePrefix={arXiv},
      primaryClass={cs.IR},
      url={https://arxiv.org/abs/2602.23766}, 
}

@inproceedings{Prometheus,
  author       = {Seungone Kim and
                  Jamin Shin and
                  Yejin Choi and
                  Joel Jang and
                  Shayne Longpre and
                  Hwaran Lee and
                  Sangdoo Yun and
                  Seongjin Shin and
                  Sungdong Kim and
                  James Thorne and
                  Minjoon Seo},
  title        = {Prometheus: Inducing Fine-Grained Evaluation Capability in Language
                  Models},
  booktitle    = {The Twelfth International Conference on Learning Representations,
                  {ICLR} 2024, Vienna, Austria, May 7-11, 2024},
  publisher    = {OpenReview.net},
  year         = {2024},
  url          = {https://openreview.net/forum?id=8euJaTveKw},
  bibsource    = {dblp computer science bibliography, https://dblp.org}
}
\newpage
\appendix
\section{Benchmark and Baseline Details}
\subsection{Details of Baselines}
The specific descriptions of compared representative baselines are as follows:

\textbf{BGE-Reranker} \cite{BGE} is a lightweight multilingual cross-encoder built on BGE-M3. It jointly encodes each query--passage pair and directly predicts its relevance, benefiting from BGE-M3's long-context and multi-granularity representations.

\textbf{GTE-Reranker} \cite{GTE} is a multilingual cross-encoder trained with contrastive relevance supervision. Its native 8,192-token encoder incorporates RoPE and unpadding to support efficient reranking of long passages.

\textbf{OS-Reranker} \cite{OpenScholar} is a scientific-domain cross-encoder used in OpenScholar. It fine-tunes a BGE reranker on synthetic query--passage relevance labels generated from scientific literature, enabling domain-specific passage prioritization.

\textbf{RankZephyr} \cite{RankZephyr} is an open-source 7B LLM-based listwise reranker that generates permutations of candidate passages. It distills ranking preferences from proprietary LLMs and employs variable-window training to improve robustness to candidate order and list size.

\textbf{CoRank} \cite{CoRank} is a training-free, model-agnostic framework for scientific literature. It first reranks a broad candidate pool using compact document features, such as categories, sections, and keywords, and then refines the leading candidates using their full texts.

\textbf{RECOMP} \cite{Recomp} compresses retrieved documents before generation using learned extractive or abstractive compressors. It selects useful sentences or synthesizes concise summaries while permitting empty outputs when the retrieved content provides no useful augmentation.

\textbf{EXIT} \cite{EXIT} performs context-aware extractive compression by jointly classifying sentences while preserving their dependencies within retrieved documents. Its parallel sentence selection adapts the retained context to both query complexity and retrieval quality.

\textbf{ChainRAG} \cite{ChainRAG} progressively rewrites decomposed sub-questions to recover missing entities and retrieves relevant sentences from a sentence graph. The sentences collected across successive steps are combined to construct the final generation context.

\textbf{SentGraph} \cite{SentGraph} constructs a hierarchical sentence graph using rhetorical nucleus--satellite relations, topic-level subgraphs, and cross-document entity bridges. It subsequently performs graph-guided sentence selection and path expansion to retrieve connected information units.

\subsection{Benchmark Statistics}
\paragraph{ScholarQABench.}

\begin{table*}[t]
  \small
  \centering
  \setlength{\tabcolsep}{5pt}
  \renewcommand{\arraystretch}{1.2}
  \resizebox{\textwidth}{!}{
  \begin{tabular}{@{}ccccc@{}}
    \toprule
    \textbf{Dataset} & \textbf{Task Format} & \textbf{Discipline}
    & \textbf{Size} & \textbf{Evaluation} \\
    \midrule
    SciFact
    & Claim $\rightarrow$ Label (True or False)
    & Biomedicine & 208 & Acc, Cite \\
    \midrule
    PubMedQA
    & Question $\rightarrow$ Answer (Yes, No)
    & Biomedicine & 843 & Acc, Cite \\
    \midrule
    QASA
    & Question $\rightarrow$ Answer (Long-form)
    & Computer Science & 1,375 & ROUGE-L, Cite \\
    \specialrule{\heavyrulewidth}{0pt}{1.2pt}
    \specialrule{\lightrulewidth}{0pt}{2pt}
    \textsc{ScholarQA-CS}
    & Question $\rightarrow$ Answer (Long-form)
    & Computer Science & 100 & Rub-Acc, Rub-Score, Cite \\
    \midrule
    \textsc{ScholarQA-Bio}
    & Question $\rightarrow$ Answer (Long-form)
    & Biomedicine & 1,451 & Rub-Score, Cite \\
    \midrule
    \textsc{ScholarQA-Neuro}
    & Question $\rightarrow$ Answer (Long-form)
    & Neuroscience & 1,308 & Rub-Score, Cite \\
    \midrule
    \textsc{ScholarQA-Multi}
    & Question $\rightarrow$ Answer (Long-form)
    & Computer Science, Physics, Biomedicine
    & 108 & Rub-Score, Cite \\
    \bottomrule
  \end{tabular}
  }
  \caption{Dataset statistics and evaluation settings of ScholarQABench.}
  \label{tab:scholarqabench}
\end{table*}

\begin{table*}[t]
  \small
  \centering
  \setlength{\tabcolsep}{3pt}
  \renewcommand{\arraystretch}{1.2}
  \begin{tabular}{@{}cccccccc@{}}
    \toprule
    \textbf{Dataset} &
    \textbf{Queries} &
    \textbf{Label Type} &
    \textbf{Reference Source} &
    \textbf{Provided Ctxs.} &
    \textbf{Avg./Query} &
    \textbf{Inserted Ctxs.} &
    \textbf{Candidate Pool} \\
    \midrule
    SciFact
    & 208 & Gold & Provided gold contexts
    & 208 & 1.00 & 179&50 \\
    \midrule
    PubMedQA
    & 843 & Gold & Provided gold contexts
    & 843 & 1.00 & 840&50 \\
    \midrule
    QASA
    & 1,375 & Gold & Indexed original contexts
    & 2,043 & 1.49 & 1,275&50 \\
    \midrule
    \textsc{ScholarQA-CS}
    & 100 & Pseudo-gold & Expert source snippets
    & 400 & 4.00 & 400 &50\\
    \midrule
    \textsc{ScholarQA-Multi}
    & 108 & Pseudo-gold & Curated input contexts
    & 647 & 5.99 & 495 &50 \\
    \bottomrule
  \end{tabular}
  \caption{Statistics of ScholarQABench-Rerank. Each query contains 50 candidates. Provided Ctxs. denotes the gold or curated source contexts included within the original ScholarQABench task instances, while Inserted Ctxs. denotes the subset absent from the initial top-50 retrieval and subsequently inserted into the candidate pool. }
  \label{tab:scholarqabench-rerank}
\end{table*}

As summarized in Table~\ref{tab:scholarqabench}, ScholarQABench combines three adapted single-paper tasks with four expert-curated multi-paper tasks \cite{OpenScholar}. SciFact reformulates biomedical claim verification as binary classification, PubMedQA evaluates yes/no biomedical question answering, and QASA requires long-form answers about computer science papers. The multi-paper subsets cover rubric-guided computer science synthesis in ScholarQA-CS, broad literature-review questions in biomedicine and neuroscience in ScholarQA-Bio and ScholarQA-Neuro, and expert-written questions and citation-grounded answers across computer science, biomedicine, and physics in ScholarQA-Multi. Together, they span classification, short-form QA, and open-ended synthesis across four scientific disciplines for comprehensive evaluation.

Table~\ref{tab:scholarqabench} also reports metrics of each task for evaluating both task-specific answer and citation quality. For SciFact and PubMedQA, accuracy measures the proportion of predictions matching the reference labels:
\[
\mathrm{Acc}
=
\frac{1}{n}
\sum_{i=1}^{n}
\mathbb{I}[\hat{y}_i=y_i].
\]
For QASA, ROUGE-L evaluates long-form answer overlap using the longest common subsequence (LCS):
\[
\mathrm{ROUGE\text{-}L}
=
\frac{2P_LR_L}{P_L+R_L},
\]
where $P_L=\mathrm{LCS}(\hat{y},y)/|\hat{y}|$ and
$R_L=\mathrm{LCS}(\hat{y},y)/|y|$.
ScholarQA-CS is additionally evaluated using expert-authored rubrics. Its rubric accuracy combines a general criterion $G$, covering answer quality and citation use, with a rubric-driven criterion $A$ based on question-specific key ingredients:
\[
\mathrm{Rub\text{-}Acc}
=
0.4G+0.6A,
\]
where must-have ingredients receive twice the weight of nice-to-have ingredients. For open-ended synthesis, Prometheus assigns scores to coverage, relevance, and organization, whose mean forms the overall rubric score:
\[
\mathrm{Rub\text{-}Score}
=
\frac{\mathrm{Cov}+\mathrm{Rel}+\mathrm{Org}}{3}.
\]
Finally, citation quality is measured on all tasks. Citation precision $P_c$ evaluates whether cited passages support the associated statements, while citation recall $R_c$ measures whether citation-worthy statements receive sufficient support. Their harmonic mean is
\[
\mathrm{Cite\text{-}F1}
=
\frac{2P_cR_c}{P_c+R_c}.
\]

\paragraph{ScholarQABench-Rerank.}
Following the same retrieval settings from OpenScholar \cite{OpenScholar}, we construct \textit{ScholarQABench-Rerank} from the five ScholarQABench tasks with gold or curated source contexts. For each query, OS-Retriever retrieves 50 passages from OS-DataStore. PubMedQA and SciFact use their provided gold contexts, while QASA identifies gold contexts through its annotated context indices. For ScholarQA-CS and ScholarQA-Multi, the expert-provided source contexts are treated as pseudo-gold, respectively. After normalization and deduplication, a candidate is labeled positive if it exactly matches, contains, or achieves a token-level F1 of at least 0.8 with a reference context. Reference contexts missing from the initial retrieval are added to the pool, and lower-ranked passages are removed to retain exactly 50 candidates per query. ScholarQA-Bio and ScholarQA-Neuro are excluded because they only contain questions and do not provide reference contexts for reranking evaluation. Table~\ref{tab:scholarqabench-rerank} summarizes the resulting benchmark.

We evaluate the ranked candidate lists using Recall@$k$, MAP@$k$, and
nDCG@$k$. Let $g_{q,i}\in\{0,1\}$ denote the gold label at rank $i$ for query
$q$, and let $G_q=\sum_{i=1}^{50}g_{q,i}$ be the number of positive candidates
in its pool. Recall@$k$ measures the proportion of positive candidates
recovered within the first $k$ positions:
\[
\mathrm{Recall}_q@k
=
\frac{\sum_{i=1}^{k}g_{q,i}}{G_q}.
\]
Average precision rewards placing positive candidates at higher ranks:
\[
\mathrm{AP}_q@k
=
\frac{
\sum_{i=1}^{k}
g_{q,i}\,\mathrm{P}_q(i)
}{
\min(G_q,k)
},
\quad
\mathrm{P}_q(i)
=
\frac{\sum_{j=1}^{i}g_{q,j}}{i}.
\]
MAP@$k$ is the mean of $\mathrm{AP}_q@k$ over all queries. Finally,
nDCG@$k$ evaluates ranking quality with logarithmic position discount:
\[
\mathrm{nDCG}_q@k
=
\frac{\mathrm{DCG}_q@k}{\mathrm{IDCG}_q@k},
\quad
\mathrm{DCG}_q@k
=
\sum_{i=1}^{k}
\frac{g_{q,i}}{\log_2(i+1)},
\]
where $\mathrm{IDCG}_q@k$ is the DCG of the ideal ordering. All metrics are
macro-averaged across queries, reported as percentages, and evaluated at
$k=10$ in our experiments.

\section{Training Data Construction}
\subsection{Construction Pipeline}
\paragraph{QA data extraction.}
We construct the training data from Open Scholar Training Data \cite{OpenScholar}, which contains 130,135
scientific instruction-tuning instances. We retain the
\texttt{single\_paper\_qa} and \texttt{multipaper\_qa} subsets and parse each
instance into a question, an answer, and a list of candidate passages. For
each passage, its reference index, title, and text are extracted from the
reference section of the original input. This process yields 25,915 QA
instances for subsequent annotation.

\paragraph{LLM-based relevance annotation.}
We use Qwen3-Max API to produce two complementary forms of
ranking supervision. At the passage level, one annotation request is created
for each QA instance, where the model ranks all candidate passages according
to their contribution to answering the question and supporting the complete
answer. At the sentence level, one request is created for every candidate
passage. The model first divides the passage into sentence-like units and then
assigns each unit a relevance rank while preserving its original textual
position. The answer is supplied to sentence annotation only for single-paper
instances containing exactly one cited passage; otherwise, sentence relevance
is judged from the question and passage alone. The complete
prompt templates are provided in Tables~\ref{tab:sentence-ranking-prompt}
and~\ref{tab:passage-ranking-prompt}, where the annotation rationales and summaries are retained only for quality examination and excluded from model training.

\paragraph{Ranking consolidation.}
The model outputs are parsed as structured JSON and aligned with their
original QA instances. Sentence units marked as fragments or containing no
meaningful content are removed. The remaining sentence ranks are converted
into strict consecutive rankings by sorting first by the annotated rank and
then by original position. Passage rankings are similarly normalized using
the annotated rank, relevance score, and original position as successive
criteria. This deterministic normalization resolves ties while preserving the
LLM-produced preference order.

\paragraph{Role enrichment.} Each retained sentence is processed by a pretrained scientific facet
model UniFAR \cite{UniFAR} to obtain the soft distribution of its semantic role over background, method, and result.
Long passages are divided into token-bounded windows before role annotation so
that every sentence receives a role distribution without truncation. The
resulting training record contains the question, passage rankings, sentence
texts and rankings, and sentence-level role distributions, providing the
supervision required by both the local scorer and the global graph reranker.

\subsection{Training Data Statistics and Validation}
\paragraph{Statistics within the pipeline.}
Table~\ref{tab:training-data-statistics} summarizes the data retained at each stage of the construction pipeline. We collect 25,915 QA instances, including 14,098 single-paper and 11,817 multi-paper examples, and obtain 257,546 question-passage pairs. Each QA instance produces one passage-ranking request, while every question-passage pair is processed by the sentence-ranking annotator. These annotations yield 2,331,189 sentence units before filtering. We remove 192,264 units marked as fragments by the LLM and 21 additional short units, retaining 2,138,904 sentences in 256,896 non-empty passages.

\paragraph{Automatic validation.}
We validate the final data at the instance, passage, and sentence levels.
Required fields are checked for every instance, and passage and sentence ranks
must each form a complete permutation from 1 to the corresponding candidate
count. We further require every retained passage to contain at least one valid
sentence and every sentence to contain both raw role attention and a normalized
three-dimensional role distribution. The final data contain no missing
rank, empty passage, malformed rank sequence, missing role field, or
non-normalized role distribution. The hard labels induced by the largest role probability comprise 35.46\% background, 32.21\% method, and 32.33\% result sentences, indicating that role annotation does not collapse to a single category.

\begin{table}[!t]
  \centering
  \small
  \setlength{\tabcolsep}{6pt}
  \begin{tabular}{lr}
  \toprule
  \textbf{Item} & \textbf{Count} \\
  \midrule
  QA instances & 25,915 \\
  \quad single-paper / multi-paper & 14,098 / 11,817 \\
  Raw question--passage pairs & 257,546 \\
  Passage-ranking requests & 25,915 \\
  Sentence-ranking requests & 257,546 \\
  Sentence units before filtering & 2,331,189 \\
  LLM-marked fragments removed & 192,264 \\
  Additional short units removed & 21 \\
  Retained sentences & 2,138,904 \\
  Passages with retained sentences & 256,896 \\
  \bottomrule
  \end{tabular}
  \caption{Training data statistics of each stage in the construction pipeline.}
  \label{tab:training-data-statistics}
\end{table}

\paragraph{Manual quality examination.}

\begin{table}[t]
  \centering
  \small
  \setlength{\tabcolsep}{8pt}
  \begin{tabular}{@{}lrr@{}}
    \toprule
    \textbf{Criterion} & \textbf{Accepted} & \textbf{Rate} \\
    \midrule
    Sentence segmentation & 2488 / 2488 & 100.00\% \\
    Sentence ranking      & 29 / 30 & 96.67\% \\
    Passage ranking       & 14 / 15 & 93.33\% \\
    Role assignment       & 43 / 45 & 95.56\% \\
    \bottomrule
  \end{tabular}
  \caption{Results of the manual quality check. Accepted instances are determined after resolving disagreements between the two annotators.}
  \label{tab:manual-quality-check}
\end{table}

\begin{table*}[t]
  \centering
  \small
  \setlength{\tabcolsep}{2pt}
  \renewcommand{\arraystretch}{1}
  \resizebox{\textwidth}{!}{
  \begin{tabular}{@{}c*{14}{c}@{}}
    \toprule
    \multirow{2}{*}{\textbf{Dataset}}
    & \multicolumn{6}{c}{\textbf{Single-Paper}}
    & \multicolumn{8}{c}{\textbf{Multi-Paper}} \\
    \cmidrule(lr){2-7}\cmidrule(l){8-15}
    & \multicolumn{2}{c}{PubMedQA}
    & \multicolumn{2}{c}{SciFact}
    & \multicolumn{2}{c}{QASA}
    & \multicolumn{2}{c}{ScholarQA-CS}
    & \multicolumn{2}{c}{ScholarQA-Multi}
    & \multicolumn{2}{c}{ScholarQA-Bio}
    & \multicolumn{2}{c}{ScholarQA-Neuro} \\
    
    \specialrule{\heavyrulewidth}{3pt}{3pt}
    
    \textit{Reranking}
    & MAP{\scriptsize @10} & nDCG{\scriptsize @10}
    & MAP{\scriptsize @10} & nDCG{\scriptsize @10}
    & MAP{\scriptsize @10} & nDCG{\scriptsize @10}
    & MAP{\scriptsize @10} & nDCG{\scriptsize @10}
    & MAP{\scriptsize @10} & nDCG{\scriptsize @10}
    & -- & -- & -- & -- \\
    \cmidrule(lr){2-3}\cmidrule(lr){4-5}
    \cmidrule(lr){6-7}\cmidrule(lr){8-9}
    \cmidrule(lr){10-11}
    seed=13
    & 84.73 & 88.83
    & 67.85 & 76.14
    & 63.98 & 66.81
    & 89.76 & 92.04
    & 30.46 & 41.29
    & -- & -- & -- & -- \\
    seed=42
    & 84.25 & 87.69
    & 67.91 & 76.33
    & 64.12 & 67.23
    & 89.68 & 92.02
    & 30.51 & 41.42
    & -- & -- & -- & -- \\
    seed=2027
    & 84.56 & 88.12
    & 67.54 & 75.88
    & 63.77 & 66.50
    & 89.55 & 91.80
    & 30.23 & 41.07
    & -- & -- & -- & -- \\
    \cmidrule(lr){2-11}
    Avg.
    & 84.51 & 88.21
    & 67.77 & 76.12
    & 63.96 & 66.85
    & 89.66 & 91.95
    & 30.40 & 41.26
    & -- & -- & -- & -- \\
    Std.
    & $\pm 0.20$ & $\pm 0.47$
    & $\pm 0.16$ & $\pm 0.18$
    & $\pm 0.14$ & $\pm 0.30$
    & $\pm 0.09$ & $\pm 0.11$
    & $\pm 0.12$ & $\pm 0.14$
    & -- & -- & -- & -- \\

    \specialrule{\heavyrulewidth}{3pt}{3pt}

    \textit{Generation}
    & Acc & Cite
    & Acc & Cite
    & R-L & Cite
    & Rub-Acc & Cite
    & Rub-Score & Cite
    & Rub-Score & Cite
    & Rub-Score & Cite \\
    \cmidrule(lr){2-3}\cmidrule(lr){4-5}
    \cmidrule(lr){6-7}\cmidrule(lr){8-9}
    \cmidrule(lr){10-11}
    \cmidrule(lr){12-13}\cmidrule(l){14-15}
    seed=13
    & 84.67 & 77.92
    & 83.17 & 55.10
    & 20.38 & 47.76
    & 62.44 & 39.42
    & 4.006 & 52.74
    & 4.240 & 54.87
    & 4.152 & 53.80 \\
    seed=42
    & 84.48 & 77.58
    & 82.96 & 54.89
    & 20.32 & 47.80
    & 62.15 & 39.21
    & 4.028 & 52.83
    & 4.256 & 55.11
    & 4.141 & 53.72 \\
    seed=2027
    & 84.39 & 77.13
    & 82.88 & 54.74
    & 19.87 & 47.54
    & 62.50 & 39.53
    & 4.003 & 52.67
    & 4.228 & 54.63
    & 4.133 & 53.46 \\
    \cmidrule(l){2-15}
    Avg.
    & 84.51 & 77.54
    & 83.00 & 54.91
    & 20.19 & 47.70
    & 62.36 & 39.39
    & 4.012 & 52.75
    & 4.241 & 54.87
    & 4.142 & 53.66 \\
    Std.
    & $\pm 0.12$ & $\pm 0.32$
    & $\pm 0.12$ & $\pm 0.15$
    & $\pm 0.23$ & $\pm 0.11$
    & $\pm 0.15$ & $\pm 0.13$
    & $\pm 0.011$ & $\pm 0.07$
    & $\pm 0.011$ & $\pm 0.20$
    & $\pm 0.008$ & $\pm 0.15$ \\
    \bottomrule
  \end{tabular}
  }
  \caption{Reranking and generation performance of LoG-Reranker across three
  random seeds. Avg. and Std. denote the mean and standard deviation over the three runs, respectively.}
  \label{tab:multi-seed-results}
\end{table*}

We additionally conduct a manual quality check by sampling 15 single-paper and 15
multi-paper instances and inspect passages and sentences drawn from the top, middle, and bottom ranking regions. Two human annotators independently evaluate four
criteria: (1) whether sentence segmentation is faithful to the source and semantically complete; (2) whether a higher-ranked sentence is more useful to the question than a lower-ranked one; (3) whether passage rankings prioritize central and, for multi-paper questions, complementary information; and (4) whether the highest-probability role agrees with the sentence's semantic function.
For sentence segmentation, we examine all 2,488 sentence units contained in
the 30 sampled instances for source fidelity and semantic completeness.
Sentence-ranking quality is assessed once for each instance by comparing
sentences from different ranking regions, resulting in 30 instance-level
judgments. Passage ranking is examined on the 15 multi-paper instances, where
the annotators consider both relevance to the central information need and
complementarity among highly ranked passages. Role assignment is evaluated on
45 sentences, comprising 15 examples sampled from each predicted role:
background, method, and result.

As shown in Table~\ref{tab:manual-quality-check}, all 2,488 inspected sentence
units preserve faithful and semantically complete content from their source
passages. Sentence rankings are accepted in 29 of the 30 instances, indicating
that the annotations reliably prioritize more useful sentences across
different ranking regions. Passage rankings satisfy the relevance and
complementarity criteria in 14 of the 15 multi-paper instances. The slightly
lower rate reflects the greater difficulty of prioritizing information across
multiple papers while avoiding redundant content. Finally, 43 of the 45
examined role assignments agree with the semantic functions of their
sentences. The two disagreements involve sentences with overlapping
functions, further supporting the use of soft role distributions. Overall,
the results indicate that the constructed data provide reliable supervision
for both fine-grained sentence scoring and cross-passage reranking.

\section{Implementation and Hyperparameters}
\subsection{Hyperparameter Sensitivity}
Table~\ref{tab:hyperparameter-sensitivity} examines the sensitivity of
LoG-Reranker to the role supervision weight and three graph-construction
parameters. Performance remains relatively stable around the default
configuration on both ScholarQA-CS and ScholarQA-Multi, indicating that the
model does not depend on a narrowly tuned parameter setting. A moderate
$\lambda_{\mathrm{role}}$ provides the best balance between sentence-role
supervision and the primary ranking objective, whereas excessively weak or
strong role supervision reduces ranking quality. For graph construction,
lower similarity thresholds and larger local-score thresholds introduce
less constrained cross-passage connections, while overly restrictive values
produce sparse graphs with limited relational information. Similarly,
retaining too few similarity neighbors restricts cross-passage propagation,
whereas a large neighborhood introduces redundant or weakly related
sentences. Although $k=16$ yields marginal gains, the improvement is at most
0.07 points while doubling the neighbor limit. Overall, the default configuration therefore provides an effective balance between ranking quality, graph connectivity, and computational efficiency.

\begin{table}[t!]
  \centering
  \small
  \setlength{\tabcolsep}{2pt}
  \begin{tabular}{@{}cccccc@{}}
    \toprule
    \multirow{2}{*}{\shortstack{\textbf{Hyper-}\\\textbf{parameter}}}
    & \multirow{2}{*}{\textbf{Value}}
    & \multicolumn{2}{c}{\textbf{ScholarQA-CS}}
    & \multicolumn{2}{c}{\textbf{ScholarQA-Multi}} \\
    \cmidrule(lr){3-4}\cmidrule(l){5-6}
    & & MAP{\scriptsize @10} & nDCG{\scriptsize @10} & MAP{\scriptsize @10} & nDCG{\scriptsize @10} \\
    \midrule

    Default
    & -- & 89.76 & 92.04 & 30.46 & 41.29 \\

    \midrule
    \multirow{4}{*}{\shortstack{Role loss\\weight $\lambda_{\mathrm{role}}$}}
    & 0.10 & 88.22 & 91.13 & 29.28 & 40.44\\
    & 0.25 & 88.79 & 91.25 & 30.10 & 40.81\\
    & 0.75 & 89.53 & 91.81 & 30.51 & 41.12\\
    & 1.00 & 89.68 & 91.57 & 30.45 & 41.33\\

    \midrule
    \multirow{4}{*}{\shortstack{Similarity\\threshold $\tau_{\mathrm{sim}}$}}
    & 0.65 & 89.12 & 90.99 & 30.19 & 40.96\\
    & 0.70 & 89.34 & 91.26 & 30.34 & 41.03\\
    & 0.80 & 89.77 & 91.95 & 30.49 & 41.18\\
    & 0.85 & 89.72 & 91.97 & 30.45 & 41.22\\

    \midrule
    \multirow{4}{*}{\shortstack{Local-score\\threshold $\delta$}}
    & 0.15 & 89.87 & 92.00 & 30.45 & 41.19\\
    & 0.20 & 89.66 & 91.89 & 30.43 & 41.21\\
    & 0.30 & 89.12 & 91.64 & 30.08 & 40.87\\
    & 0.35 & 88.75 & 91.43 & 29.91 & 40.10\\

    \midrule
    \multirow{4}{*}{\shortstack{Similarity\\neighbors $k$}}
    & 2  & 87.98 & 90.55 & 29.01 & 39.86\\
    & 4  & 88.74 & 91.23 & 30.17 & 40.72\\
    & 12 & 89.78 & 92.02 & 30.51 & 41.26\\
    & 16 & 89.83 & 92.10 & 30.53 & 41.33\\

    \bottomrule
  \end{tabular}
  \caption{Hyperparameter sensitivity on ScholarQA-CS and ScholarQA-Multi.
  Each parameter is varied independently while the others remain at their
  default configuration: $\lambda_{\mathrm{role}}=0.5$,
  $\tau_{\mathrm{sim}}=0.75$, $\delta=0.25$, and $k=8$.}
  \label{tab:hyperparameter-sensitivity}
\end{table}

\subsection{Random Seeds}
We evaluate the stability of LoG-Reranker using three random seeds: 13, 42,
and 2027. Table~\ref{tab:multi-seed-results} reports the individual results
together with their means and standard deviations. The main results use seed
13, while the additional runs follow the same training and evaluation
settings.

LoG-Reranker exhibits low variance across both reranking and generation
tasks. Standard deviations for reranking metrics range from 0.09 to 0.47,
with particularly stable results on ScholarQA-CS and ScholarQA-Multi. For
generation, the standard deviations are at most 0.23 for task-specific
metrics and 0.32 for citation F1, while the three multi-paper Rub-Scores vary
by no more than 0.011. The average results also remain close to those obtained
with seed 13 across all datasets. These results indicate that the improvements
of LoG-Reranker are robust to random initialization and are not driven by a
particular training run.

\section{Additional Experiments}
\subsection{Efficiency and Graph Complexity}
\label{sec:efficiency}
Table~\ref{tab:efficiency} compares the end-to-end reranking efficiency
and graph complexity of all methods on the full ScholarQA-CS dataset.
Each method reranks the same pool of 50 candidate passages per query.
Latency is the average wall-clock time per query, while memory denotes
peak local GPU memory usage. Parameter counts include all locally loaded
model components. For methods relying on remote APIs, server-side
parameters and memory consumption are not observable and are therefore
marked as unavailable.

Cross-encoders and the lightweight RECOMP model achieve
the lowest latency because they do not explicitly model relations among
sentences from different passages. In contrast, LLM-based methods incur
substantially higher computational cost. RankZephyr requires 7.24B
parameters and 15.48\,GB of GPU memory, while CoRank takes 291.335
seconds per query because of its multi-stage API-based inference.
EXIT is also computationally expensive, requiring 2.51B parameters,
15.20\,GB of memory, and 63.113 seconds per query. These results
illustrate the considerable efficiency cost of applying large
language models to passage reranking.

Among sentence graph-based methods, LoG-Reranker provides the most
efficient end-to-end inference. Its average latency is 2.426 seconds,
making it approximately $2.6\times$ faster than ChainRAG and
$40.2\times$ faster than SentGraph. Unlike these two methods,
LoG-Reranker performs reranking entirely with locally deployed models
and does not introduce remote API latency. Its complete local and global
reranking pipeline contains 152.52M parameters, comprising 149.85M
parameters in the local reranker and 2.67M in the global reranker.
This model size is close to that of GTE-Reranker and substantially
smaller than those of BGE-Reranker, OS-Reranker, RankZephyr, and EXIT.
The additional graph reasoning increases peak memory to 3.21\,GB, but
the requirement remains far below those of the large-model baselines.

LoG-Reranker also constructs a more compact but more richly connected
evidence graph. It retains 530.6 sentence nodes on average, compared
with 577.2 for both ChainRAG and SentGraph, corresponding to an
$8.1\%$ reduction. Meanwhile, its graph contains 3387.3 unique relation
pairs on average, which is $62.4\%$ more than ChainRAG and more than
four times that of SentGraph. This combination indicates that LoG does
not obtain global evidence coverage by retaining more sentences.
Instead, it forms richer connections among a smaller set of selected
evidence units. Overall, the results demonstrate that LoG-Reranker
achieves a balance between graph expressiveness and practical
inference efficiency.
\begin{table}[th!]
  \centering
  \small
  \setlength{\tabcolsep}{1.5pt}
  \resizebox{\columnwidth}{!}{
  \begin{tabular}{cccccc}
  \toprule
  \textbf{Method}
  & \textbf{Params.}
  & \textbf{Latency{\scriptsize(S)}}
  & \textbf{Memory{\scriptsize(GB)}}
  & \textbf{Nodes}
  & \textbf{Edges} \\
  \midrule
  
  \multicolumn{6}{c}{\textit{Cross-encoder}} \\
  \cmidrule(lr){1-6}
  BGE-Reranker & 567.8M & 1.092 & 2.00 & -- & -- \\
  GTE-Reranker & 149.6M & 0.524 & 1.06 & -- & -- \\
  OS-Reranker  & 559.9M & 0.651 & 1.96 & -- & -- \\
  
  \cmidrule(lr){1-6}
  \multicolumn{6}{c}{\textit{LLM-based}} \\
  \cmidrule(lr){1-6}
  RankZephyr & 7.24B & 13.538  & 15.48  & -- & -- \\
  CoRank     & API   & 291.335 & -- & -- & -- \\
  
  \cmidrule(lr){1-6}
  \multicolumn{6}{c}{\textit{Fine-grained}} \\
  \cmidrule(lr){1-6}
  RECOMP & 109.5M & 0.421  & 0.70  & -- & -- \\
  EXIT   & 2.51B  & 63.113 & 15.20 & -- & -- \\
  
  \cmidrule(lr){1-6}
  \multicolumn{6}{c}{\textit{Sentence graph-based}} \\
  \cmidrule(lr){1-6}
  ChainRAG
  & 696.7M + API & 6.208  & 2.06 & 577.2 & 2085.9 \\
  SentGraph
  & 136.8M + API & 97.629 & 0.94 & 577.2 & 780.6 \\
  \textbf{LoG-Reranker}
  & 152.5M & 2.426 & 3.21 & 530.6 & 3387.3 \\
  \bottomrule
  \end{tabular}
  }
  \caption{Efficiency and graph complexity on the ScholarQA-CS dataset with 50 candidate passages per query.}
  \label{tab:efficiency}
\end{table}

\subsection{Input Length and Generation Quality}
\label{sec:input_efficiency}
We analyze input length as an outcome of reranking rather than enforcing
an identical token budget. Different rerankers naturally produce
evidence at different granularities: passage-level methods preserve
complete passages, whereas sentence-level methods select or organize
fine-grained evidence. Imposing a fixed token budget would require
truncating the native outputs of some methods or supplementing others
with lower-ranked evidence, thereby changing the reranking decisions
being evaluated. We therefore retain the native top-10 outputs of each
method and use the same generator and generation prompt in all settings.

\begin{table*}[t!]
  \centering
  \setlength{\tabcolsep}{9pt}
  \begin{tabular}{lccccccc}
  \toprule
  \multirow{2}{*}{\textbf{Method}}
  & \multirow{2}{*}{\textbf{Level}}
  & \multicolumn{3}{c}{\textbf{ScholarQA-CS}}
  & \multicolumn{3}{c}{\textbf{ScholarQA-Multi}} \\
  \cmidrule(lr){3-5}\cmidrule(lr){6-8}
  &
  & \textbf{Input Tokens}
  & \textbf{Rub-Acc}
  & \textbf{Cite}
  & \textbf{Input Tokens}
  & \textbf{Rub-Score}
  & \textbf{Cite} \\
  \midrule
  BGE-Reranker
  & Passage  & 6761.27 & 60.86 & 36.23
  & 3239.31 & 3.910 & \underline{52.00} \\

  GTE-Reranker
  & Passage  & 6659.04 & 59.56 & 31.02
  & 3068.59 & \underline{3.975} & 49.97 \\

  OS-Reranker
  & Passage  & 6749.36 & \underline{61.72} & 37.08
  & 3215.95 & 3.923 & 51.71 \\

  RankZephyr
  & Passage  & 3569.38 & 58.70 & 32.68
  & 2873.06 & 3.871 & 47.24 \\

  CoRank
  & Passage  & 6636.51 & 61.15 & 35.89
  & 3001.44 & 3.964 & 50.95 \\

  RECOMP
  & Sentence & 1340.08 & 55.39 & 33.81
  & 1434.70 & 3.735 & 41.36 \\

  EXIT
  & Sentence & 881.61 & 54.80 & 37.51
  & 902.56 & 3.682 & 37.16 \\

  ChainRAG
  & Sentence & 2020.62 & 59.47 & \underline{38.76}
  & 1376.48 & 3.753 & 48.98 \\

  SentGraph
  & Sentence & 2364.23 & 61.32 & 38.14
  & 1513.68 & 3.797 & 49.23 \\

  \textbf{LoG-Reranker}
  & Sentence & 3173.11 & \textbf{62.44} & \textbf{39.42}
  & 1991.14 & \textbf{4.006} & \textbf{52.74} \\
  \bottomrule
  \end{tabular}
  \caption{Generation performance and average input length on ScholarQA-CS
  and ScholarQA-Multi.}
  \label{tab:input_efficiency}
\end{table*}

\begin{table*}[t!]
  \centering
  \setlength{\tabcolsep}{6.5pt}
  \begin{tabular}{@{}ccccccccc@{}}
  \toprule
  \multirow{2}{*}{} & \multicolumn{4}{c}{ScholarQA-CS} & \multicolumn{4}{c}{ScholarQA-Multi} \\ \cmidrule(l){2-5} \cmidrule(l){6-9} 
   & MAP{\scriptsize @10} & nDCG{\scriptsize @10} & Rub-Acc & Cite & MAP{\scriptsize @10} & nDCG{\scriptsize @10} & Rub-Score & Cite \\ \midrule
  \textbf{Full LoG-Reranker} & \textbf{89.76} & \textbf{92.04} & \textbf{62.44} & \textbf{39.42} & \textbf{30.46} & \textbf{41.29} & \textbf{4.006} & \textbf{52.74} \\ \midrule
  w/o contextual & 87.12 & 89.46 & 59.37 & 38.59 & 28.25 & 38.78 & 3.853 & 51.67 \\
  w/o role-head & 87.96 & 90.43 & 61.23 & 38.87 & 29.02 & 39.86 & 3.882 & 51.88 \\
  w/o intent-head & 88.54 & 91.08 & 62.12 & 39.21 & 29.85 & 40.77 & 3.958 & 52.23 \\ \midrule
  w/o intra-passage edges & 87.54 & 90.15 & 60.01 & 38.66 & 28.79 & 39.14 & 3.867 & 51.93 \\
  w/o cross-passage edges & 87.15 & 89.23 & 59.38 & 38.13 & 28.24 & 38.65 & 3.711 & 51.16 \\
  w/o edge constraints & 89.03 & 91.26 & 61.85 & 39.04 & 29.41 & 40.82 & 3.876 & 52.06 \\
  w/o edge types & 88.24 & 90.57 & 61.09 & 38.37 & 28.93 & 40.15 & 3.788 & 51.93 \\ \midrule
  w/o adjacency expansion & -- & -- & 62.23 & 39.50 & -- & -- & 3.895 & 52.65 \\
  w/o similarity expansion & -- & -- & 62.31 & 39.39 & -- & -- & 3.912 & 52.72 \\ \bottomrule
  \end{tabular}
  \caption{Detailed ablation results on ScholarQA-CS and ScholarQA-Multi.Dashes indicate
variants that modify only the structured generation context and leave reranking results unchanged.}
  \label{tab:detail-ablations}
\end{table*}

Table~\ref{tab:input_efficiency} shows that generation quality is not
determined simply by providing more tokens. On both datasets,
LoG-Reranker uses fewer input tokens than every passage-level baseline
while achieving the best generation and citation results. In
ScholarQA-CS, LoG obtains a Rub-Acc of 62.44 and a citation score of 39.42
with 3173 input tokens, compared with approximately 6600--6800 tokens
for the strongest full-passage rerankers. The same trend holds in
ScholarQA-Multi, where LoG achieves the highest Rub-Score of 4.006 and
citation score of 52.74 using only 1991 tokens.

Among sentence-level methods, very short inputs do not necessarily lead
to strong generation. RECOMP and EXIT substantially reduce input length
but also lose important supporting evidence, resulting in lower rubric
and citation scores. LoG instead enriches selected main sentences with
targeted local and cross-passage evidence. This produces longer inputs
than aggressive sentence compressors but remains considerably more
compact than passage-level reranking. The consistent improvements
across all four generation metrics suggest that LoG selects and
organizes more task-critical evidence, achieving a better balance
between context compactness and evidence completeness.

\subsection{Detailed Ablation Results}

Table~\ref{tab:detail-ablations} further isolates the contributions of
individual components within the local scoring, global graph reranking, and
structured context stages.

\textbf{Local sentence modeling.}
Removing contextual encoding causes the largest degradation among the local
variants, reducing Rub-Acc by 3.07 points on ScholarQA-CS and Rub-Score by
0.153 on ScholarQA-Multi. This confirms that sentence relevance cannot be
estimated reliably without preserving its passage context. Removing either
the role or intent head also consistently reduces ranking and generation
quality, with the larger decline from removing the role head indicating that
sentence function provides a particularly important signal for fine-grained
selection.

\textbf{Graph structure and relation modeling.}
Cross-passage edges provide the strongest contribution within the global
stage. Their removal decreases nDCG@10 by 2.81 and 2.64 points on the two
tasks and produces the lowest ScholarQA-Multi Rub-Score of 3.711. In-passage
edges are likewise important for retaining local continuity. The declines
caused by removing edge constraints or relation types further show that graph
quality depends not only on adding connections, but also on filtering and
distinguishing their relational semantics.

\textbf{Structured context expansion.}
Removing adjacency or similarity expansion does not alter sentence rankings
but lowers the task-specific generation scores. Adjacency expansion yields
the larger benefit, particularly on ScholarQA-Multi, suggesting that restoring
local sentence context is important after fine-grained selection. Citation F1
changes only marginally for these variants, indicating that neighbor expansion
primarily improves the completeness and organization of the synthesis rather
than source attribution.

\begin{table*}[p]
  \centering
  \small
  \setlength{\tabcolsep}{8pt}
  \renewcommand{\arraystretch}{1.0}
  \begin{tabular}{p{\dimexpr\textwidth-2\tabcolsep\relax}}
    \toprule
    \textbf{System Message} \\
    \midrule
    You are a careful scientific data annotator. Return valid JSON only. \\
    \midrule
    \textbf{User Prompt} \\
    \midrule

    Task: sentence-level relevance ranking inside one scientific passage.
    \par\smallskip
    Given a question and one candidate passage, split the passage text into
    sentence-like units, preserve their original order, and assign each unit a
    relevance rank for answering the question.
    \par\smallskip
    Ranking rules:
    \par
    - Rank 1 is the most useful sentence for answering the question.
    \par
    - Rank by answerability and evidence utility, not by broad
    topical similarity or lexical overlap alone.
    \par
    - Prioritize evidence in this order:
    \par
    \hspace{1em}1. Sentences that explicitly answer the question or state the
    key contrast/relation.
    \par
    \hspace{1em}2. Sentences that explain the specific mechanisms, dimensions,
    variables, or evidence needed to understand that answer.
    \par
    \hspace{1em}3. Sentences that establish the compared entities, task
    setting, or study object.
    \par
    \hspace{1em}4. Weak conceptual interpretation or paper-level significance
    that mentions answer-related concepts but does not state the answer.
    \par
    \hspace{1em}5. Generic experimental methods, instrument lists, irrelevant
    background, or incomplete fragments.
    \par
    - Do not over-rank meta-commentary such as novelty, importance,
    or broad interpretation unless it directly supports answering the question.
    \par
    - Rationales must describe the evidence semantics in the
    passage. Do not say or imply that a sentence is important because it
    appears in, matches, or is cited by the gold answer.
    \par
    - Pure method or instrumentation sentences should be ranked
    very low unless they themselves identify answer-specific mechanisms,
    variables, or dimensions.
    \par
    - For overview, review, survey, or current-state questions, rank
    sentence-like units higher when they state representative findings, major
    trends, categories, limitations, debates, or future directions, rather
    than isolated details.
    \par
    - Use consecutive integer ranks from 1 to N, where N is the
    number of sentence-like units you output.
    \par
    - Output sentence-like units in their original passage order,
    not ranked order. This is mandatory: do not sort the output by rank.
    \par
    - Keep each unit faithful to the source text. Do not invent
    content.
    \par
    - If a sentence contains separable clauses with clearly
    different evidence value, you may split it into smaller sentence-like
    units, but preserve the original text order.
    \par
    - If a passage has bullet-like fragments or formulas, group
    them into readable sentence-like units.
    \par
    - If a sentence-like unit is an incomplete fragment caused by
    passage truncation or random splitting, keep it and set
    \texttt{is\_fragment=true}. Otherwise set
    \texttt{is\_fragment=false}.
    \par
    - Assign \texttt{relevance\_score} after deciding the rank.
    Prefer differentiated scores for short passages:
    \par
    \hspace{1em}5 = direct answer / decisive evidence;
    \par
    \hspace{1em}4 = key supporting mechanism or explanatory evidence;
    \par
    \hspace{1em}3 = useful context that establishes entities or setting;
    \par
    \hspace{1em}2 = weak conceptual/meta interpretation or indirect context
    that still mentions answer-related concepts;
    \par
    \hspace{1em}1 = generic method-only/instrumentation, irrelevant
    background, or incomplete fragment.
    \par
    - Scores should be broadly consistent with ranks: lower-ranked
    units should not receive stronger scores than higher-ranked units unless
    the rationale explains a near-tie.
    \par\smallskip
    Question:
    \texttt{\{question\}}
    \par\smallskip
    Gold answer is provided only as auxiliary evidence for this cited passage.
    Use it to understand what information is answer-supporting, but do not copy
    it.
    \par
    Gold answer:
    \texttt{\{answer\}}
    \par\smallskip
    Passage index: \texttt{\{passage\_index\}}
    \par
    Passage title: \texttt{\{passage\_title\}}
    \par
    Passage text:
    \texttt{\{passage\_text\}}
    \par\smallskip
    Return strict JSON only with this schema:
    \par
    \texttt{\{}
    \par
    \quad\texttt{"sentences": [}
    \par
    \qquad\texttt{\{}
    \par
    \qquad\quad\texttt{"sentence\_id": 0,}
    \par
    \qquad\quad\texttt{"sentence": "original sentence text",}
    \par
    \qquad\quad\texttt{"relevance\_rank": 1,}
    \par
    \qquad\quad\texttt{"relevance\_score": 5,}
    \par
    \qquad\quad\texttt{"is\_fragment": false,}
    \par
    \qquad\quad\texttt{"rationale": "short evidence-focused reason; do not
    mention gold answer or annotation process"}
    \par
    \qquad\texttt{\}}
    \par
    \quad\texttt{]}
    \par
    \texttt{\}}
    \\
    \bottomrule
  \end{tabular}
  \caption{Sentence-level ranking prompt used for training data construction.}
  \label{tab:sentence-ranking-prompt}
\end{table*}

\begin{table*}[p]
  \centering
  \small
  \setlength{\tabcolsep}{8pt}
  \renewcommand{\arraystretch}{1.0}
  \begin{tabular}{p{\dimexpr\textwidth-2\tabcolsep\relax}}
    \toprule
    \textbf{System Message} \\
    \midrule
    You are a careful scientific data annotator. Return valid JSON only. \\
    \midrule
    \textbf{User Prompt} \\
    \midrule

    Task: passage-level relevance ranking across candidate scientific passages.
    \par\smallskip
    Given a question, candidate passages, and a gold answer, rank the passages
    by how useful they are for answering the question and supporting the answer.
    \par\smallskip
    Ranking rules:
    \par
    - Rank 1 is the most useful passage for answering the question and
    supporting the answer.
    \par
    - Use consecutive integer ranks from 1 to N, where N is the number of
    passages.
    \par
    - Output passages in their original input order, not ranked order. This is
    mandatory: do not sort the output by rank.
    \par
    - Base your judgment on whether the passage text provides evidence needed
    to construct or support the answer.
    \par
    - Rank passages by global answer contribution, not only by independent
    topical similarity.
    \par
    - For multi-part, comparative, or synthesis questions, prefer passages that
    contribute complementary evidence covering different aspects of the
    answer, such as contrasts, mechanisms, populations, settings, limitations,
    quantitative evidence, or implications.
    \par
    - Prefer passages central to the main information need over narrow
    examples, isolated statistics, or repeated evidence for an already-covered
    aspect.
    \par
    - For comparison or contrast questions, passages that explicitly discuss
    differences, trade-offs, or contrasting conditions/settings should
    generally rank higher than passages discussing only one side in isolation.
    \par
    - For overview, review, survey, or current-state questions, prioritize
    passages that cover major themes, representative findings, trends,
    limitations, debates, or future directions. Prefer complementary coverage
    across subtopics over repeated evidence for the same narrow point.
    \par
    - Consider marginal contribution: a passage is more valuable if it adds a
    new necessary aspect to the evidence pool, not merely because it repeats
    information found elsewhere.
    \par
    - Do not invent evidence not present in the passage.
    \par
    - Evidence summaries must describe the passage semantics. Do not say or
    imply that a passage is important because it appears in, matches, or is
    cited by the gold answer.
    \par
    - Evidence summaries should briefly state: (1) what aspect of the question
    the passage supports, (2) what evidence or findings it provides, and, when
    useful, (3) how it complements other passages.
    \par
    - Assign \texttt{relevance\_score} after deciding the rank:
    \par
    \hspace{1em}5 = central evidence that directly supports major parts of the
    answer or provides indispensable comparative/synthesis information;
    \par
    \hspace{1em}4 = strong supporting evidence covering an important aspect,
    mechanism, population, setting, limitation, or implication;
    \par
    \hspace{1em}3 = partially useful evidence or context for a secondary
    aspect;
    \par
    \hspace{1em}2 = weakly related contextual/background discussion;
    \par
    \hspace{1em}1 = irrelevant, generic background, repeated low-value
    evidence, or off-topic content.
    \par
    - Scores should be broadly consistent with ranks; use the rationale to
    clarify near-ties or complementary but non-central passages.
    \par\smallskip
    Question:
    \texttt{\{question\}}
    \par\smallskip
    Gold answer:
    \texttt{\{answer\}}
    \par\smallskip
    Candidate passages:
    \texttt{\{candidate\_passages\}}
    \par\smallskip
    Return strict JSON only with this schema:
    \par
    \texttt{\{}
    \par
    \quad\texttt{"passages": [}
    \par
    \qquad\texttt{\{}
    \par
    \qquad\quad\texttt{"original\_position": 0,}
    \par
    \qquad\quad\texttt{"passage\_index": 0,}
    \par
    \qquad\quad\texttt{"relevance\_rank": 1,}
    \par
    \qquad\quad\texttt{"relevance\_score": 5,}
    \par
    \qquad\quad\texttt{"evidence\_summary": "short evidence-focused reason; do
    not mention gold answer or annotation process"}
    \par
    \qquad\texttt{\}}
    \par
    \quad\texttt{]}
    \par
    \texttt{\}}
    \\
    \bottomrule
  \end{tabular}
  \caption{Passage-level ranking prompt used for training data construction.}
  \label{tab:passage-ranking-prompt}
\end{table*}

\end{document}